\documentclass[sn-basic]{sn-jnl}

\usepackage{graphicx}%
\usepackage{multirow}%
\usepackage{amsmath,amssymb,amsfonts}%
\usepackage{amsthm}%
\usepackage{mathrsfs}%
\usepackage[title]{appendix}%
\usepackage{xcolor}%
\usepackage{textcomp}%
\usepackage{manyfoot}%
\usepackage{booktabs}%
\usepackage{algorithm}%
\usepackage{algorithmicx}%
\usepackage{algpseudocode}%
\usepackage{listings}%

\usepackage{abstract,lipsum}
\usepackage{comment}

\usepackage{natbib}
\usepackage{enumitem}
\usepackage{caption}
\usepackage{subcaption}
\usepackage{array}
\usepackage{hyperref}
\usepackage{etoolbox}
\usepackage{lmodern}
\usepackage{xcolor}
\usepackage{color, colortbl}

\usepackage{soul}
\usepackage[colorinlistoftodos,textwidth=1.8cm, textsize=tiny]{todonotes}

\begin{document}

\title{Lasso Multinomial Performance Indicators for in-play Basketball Data}

\vspace{-2ex}
\author[1,2,3]{\vspace{4ex}Argyro Damoulaki}
\author[1,2,3]{Ioannis Ntzoufras} 
\affil[1]{Department of Statistics, Athens University Economics Business, Athens, Greece}
\affil[2]{AUEB Sports Analytics Group, Computational and Bayesian Statistics Lab}
\affil[3]{Institute of Statistical Research Analysis \& Documentation (ISTAER)}
\author[4]{Konstantinos Pelechrinis}
\affil[4]{Department of Informatics and Networked Systems, University of Pittsburgh, Pittsburgh, United States}

\maketitle
\vspace{-10ex} 

\begin{abstract}
\vspace{2ex} A typical approach to quantify the contribution of each player in basketball uses the plus-minus method.  The ratings obtained by such a method are estimated using simple regression models and their regularized variants, with response variable being either the points scored or the point differences.
To capture more precisely the effect of each player, detailed possession-based play-by-play data may be used. 
This is the direction we take in this article, in which we investigate the performance of regularized adjusted plus-minus (RAPM) indicators estimated by different regularized models having as a response the number of points scored in each possession.  
Therefore, we use possession play-by-play data from all NBA games for the season 2021-22 (322,852 possessions).
We initially present simple regression model-based indices starting from the implementation of ridge regression which is the standard technique in the relevant literature.
We proceed with the lasso approach which has specific advantages and better performance than ridge regression when compared with selected objective validation criteria. 
Then, we implement regularized binary and multinomial logistic regression models to obtain more accurate performance indicators since the response is a discrete variable taking values mainly from zero to three. 
Our final proposal is an improved RAPM measure which is based on the expected points of a multinomial logistic regression model where each player's contribution is weighted by his participation in the team's possessions. 
The proposed indicator, called weighted expected points (wEPTS), outperforms all other RAPM measures we investigate in this study.
\end{abstract}
\vspace{1.5ex}
{Keywords: plus-minus, RAPM, regularization, logistic regression, expected points (EPTS)}


\section{Introduction}
In team sports, one of the most relevant and valuable metrics is the impact a player has on the outcome of his team. As the renowned Dean Oliver stated {\it``Teamwork is the element of basketball most difficult to capture in any quantitative sense''} 
\citep{oliver_2002}.
Using only the information of the players' teammates and opponents on the court, \cite{rosenbaum_2004} introduced Adjusted Plus-Minus (APM) to measure the players' contribution. Later, \cite{ilardi_barzilai_2007, ilardi_barzilai_2008} suggested a ridge regression approach to address multicollinearity issues in APM, leading to the development of Regularized Adjusted Plus-Minus (RAPM).

In this work, we study the performance of several approaches for the players' evaluation of their offensive and defensive contributions using play-by-play NBA data. 
Each data observation (row in our dataset) refers to an offence (or equivalently possession). 
The response variable of interest is the outcome (i.e. achieved score) per possession that ranges from zero to six. 
Someone may be surprised by this range since, in basketball, the natural outcomes are from zero to three points. The additional values (4--6) are rare and may occur only due to offensive goal fouls or technical fouls.

In our analysis, we use the ridge regression approach as a starting point. 
A common issue in such studies is the performance evaluation of low-time players, whose RAPMs are usually found to be very high indicating that they over-perform in comparison to their actual contribution. 
As a second step, we applied lasso regularization method instead of ridge for the normal regression model. This improved the results, as lasso distinguishes the better and worse players more reasonably and realistically than the ridge regression. Moreover, players with average or non-important RAPM are automatically set to zero from the lasso approach. We consider this as an advantage of lasso over ridge regression since we are usually interested for evaluating only the top-rated or efficient players --- in some occasions, we may also be interested in identifying the worst or inefficient players. Nevertheless, someone may argue that estimating a RAPM evaluation metric for every player is more desirable. 
A major disadvantage of both of these approaches is the use of the normal distribution  to model a discrete response (the outcome per possession). Clearly, the normal distribution is not suitable to stochastically model such a response and other alternatives should be found to build a more accurate and statistically correct model.

Hence, we extended the standard ridge regression approach by using binary and multinomial logistic regression models. We first simplified the problem by using a regularized binomial model with a response of a binary variable indicating whether the offensive team scored or not. Surprisingly, the logistic regression RAPM estimates were found to be highly correlated with the usual RAPM values. In this way, we provide a different, valid interpretation for the usual normal regression-based RAPMs.

Finally, we produce a regularized multinomial logistic regression model for the points scored per possession. The model was fitted indirectly through three separate binomial models for the different types of points scored (one, two, and 3+). Here, a new evaluation metric is proposed based on the expected points per possession (EPTS) and its extended version, the weighted expected points (wEPTS).
The latter index takes into consideration the proportion of possessions that each player participates in his team. 
It is derived as the points expected to be scored by the team over all games 
under the simplified scenario that all other players on court belong to the reference group with zero-lasso RAPM measures. 

Our initial intuition that the multinomial model is more suitable for modelling the points per possession and for producing player evaluation metrics is confirmed since the obtained wEPTS evaluation index is found to be superior according to selected external validation criteria. 
Moreover, this metric seems to partially solve the problem related to the overestimation of the performance of low-time players. 
Another characteristic of the proposed regularized multinomial approach is that the overall EPTS (and wEPTS) index can be decomposed in three different RAPM measures, one for each possible scoring outcome (one, two and 3+). Therefore, the method can spot the best-performed players according to each scoring outcome. 
For example, in the modern NBA the value of the three point shot cannot be overstated. However, there can be players that while they do not take a high volume of threes they contribute significantly to their teammates taking and making higher quality threes. For example, Giannis Antetokounmpo and LeBron James are not high volume three point shooters, but their playmaking abilities, in conjunction with their {\em gravity} allows them to generate high quality three point shots for their teammates. 
Our model will be able to identify these contributions from players and this is a unique characteristic of our approach that no other previously proposed method has. 

The structure of the paper is as follows. Section \ref{literature} presents the main background in sports and basketball analytics regarding plus-minus, Section \ref{sec2_data_methods} describes the data and the methodology used in this study. Section \ref{sec3_results} presents the results of our analysis regarding RAPM and EPTS ratings and, in Section~\ref{sec4_wEPTS}, we introduce the finally proposed weighted expected points (wEPTS) rating. Section \ref{sec5_discussion} concludes with some interesting findings and discussions.

\section{Background}
\label{literature}
Due to developments in technology and the growing availability of detailed game data, the application of advanced analytics in basketball has experienced significant growth in recent years. Initially, basketball analytics concentrated on fundamental statistical metrics, such as points per game and field goal percentage but, in recent years, more advanced and sophisticated methods have been developed.
As \cite{oliver_2002}, discusses in Basketball on Paper, understanding basketball from an analytical perspective involves considering not only individual metrics but also team dynamics and the interaction between offensive and defensive strategies.

In the context of player evaluation, the primary challenge lies in ``isolating'' the contribution of individual players from their teammates and opponents. This led to the development of various metrics designed to quantify a player’s impact on their team.
Plus-minus ratings have been a key tool in evaluating player performance in team sports, especially in basketball. Traditional plus-minus assesses a team’s net performance, typically expressed as a point differential over 100 possessions, when a specific player is on the court. However, its simplicity can be misleading; players on strong teams who face weaker opponents may appear to have excessive ratings, while elite players on underperforming teams might receive an unexpectedly low rating that does not correspond to their quality of play.
The introduction of Adjusted Plus-Minus (APM) by \cite{rosenbaum_2004} is an important advance over traditional plus-minus by accounting for both teammate and opponent quality, reducing the biases of earlier models. 
While the introduction of APM was well received by both academics and professional basketball analysts for capturing players' impact without needing to assign in-game actions or box-score statistics, it has certain limitations; (a) its instability, (b) the presence of extreme outliers and (c) collinearity of player indicators. 
Regarding instability, a player's APM can vary greatly from season to season, making this metric unreliable for systematic evaluation. Concerning the presence of extreme outliers, this refers to certain players who may exhibit unexpectedly high or low APM scores. Additionally, we have the issue of low-minute players, who appear as outliers primarily due to two reasons: the lack of playing data and the quality of their data, as they usually play during time periods that are not critical for their team (garbage time). Finally, collinearity refers to the fact that specific players often play together in the same lineup, making difficult to estimate each player's individual contribution.

\citeauthor{rosenbaum_2004}'s \citeyearpar{rosenbaum_2004}  
regression-based APMs have inspired many researchers in the field of sports analytics. 
Starting from basketball, \citet{ilardi_barzilai_2007, ilardi_barzilai_2008} implemented the model by incorporating ridge regression, a regularization technique that resolves multicollinearity issues. Ridge regression is the main tool for the development of the Regularized Adjusted Plus-Minus (RAPM) model, as it provides stable estimates, especially for players with limited playing time. RAPM’s advantages include:
i)low-variance estimates;
ii)an evaluation metric for all players;   
iii)more stable  evaluation metrics for players that frequently play together because it 
handles multicollinearity in the fitted model; and 
iv)easy interpretation of the RAPM evaluation metrics via the underlying fitted linear regression.

Another modification was proposed by \cite{mcdonald_2011_hockey}, who developed two weighted least squares regression models to estimate a National Hockey League (NHL) player's effect on his team's success in scoring and preventing goals, independent of that player's teammates and opponents. One year later, \cite{mcdonald_2012_hockey} additionally applied the ridge regularization technique in his implementation. 

A notable advancement in the field was  the development of WINVAL software system, introduced by \cite{winston_2012}. Unlike traditional methods that focus on the average point difference per possession, WINVAL evaluates the impact a player has on probability of win through the score difference when he is on the court. However, the system still faces limitations, such as over-evaluating players who consistently play alongside strong teammates or face weaker opponents at meaningless points of the game (like garbage time).

One of the main characteristics of RAPM estimated via ridge regression is that the method assigns non-zero ratings to all players. In contrast, lasso regression (Least Absolute Shrinkage and Selection Operator), introduced by \citet{tibshirani1996regression}, treats RAPM ratings differently by shrinking the ratings of specific players to zero.
In sports analytics, this regularization method is not so popular. However, \cite{Gramacy2013}, and, later, \cite{Gramacy2015} presented Bayesian lasso in logistic regression to model the probability of scoring a goal by the home team in hockey by taking advantage of the variable selection implied by lasso. 

Following the trend of model-based plus-minus ratings, weighted metrics are another alternative approach to measuring player performance.
For instance, \cite{matano_2018_soccer} noted that APM does not have the same impact in soccer as it does in basketball and hockey, since soccer is a low-scoring sport, with a low number of substitutions. 
Therefore, they developed the Augmented APM by combining FIFA ratings with APM, resulting in stronger predictive ability than standard APM or a simple regression model using only FIFA ratings.
This metric, also, considers the problem of collinearity among players. The idea is to develop APM through a Bayesian framework and incorporate FIFA ratings into the prior distribution.
\cite{schultze_2018_soccer_weighted} introduced the use of a different combination of features
in the model used for the estimation of RAPMs for individual soccer player performance called the weighted plus-minus metric. 
This metric is also based on considering the opponent team's strength (according to betting odds) and the goal's importance (considering the impact of a goal on the game's outcome). 
Finally, feature-enhanced ratings have also been popular in basketball in recent years.
They are mainly based on box-score statistics and tracking data; 
see for example 
RAPTOR 
PIE, 
DARKO 
and LEBRON; see in NBAstuffer website for a detailed list including references (\url{https://www.nbastuffer.com/analytics101/}).

Bayesian models in basketball analytics have gained popularity the recent years.
\cite{fearnhead_taylor_2010} introduced an adjusted plus-minus model within a Bayesian framework to estimate offensive and defensive ratings, assuming these ratings follow distinct Gaussian distributions. They incorporated player rating variations over time by using each season’s ratings on the prior season's results, adjusted toward a prior mean. Later, \cite{deshpande_2016} extended this idea by applying Bayesian linear regression with independent Laplacian priors for player and team components, suggesting the ratio of the posterior mean to the posterior standard deviation of the player’s partial effect to compare players.
A recent noteworthy contribution comes from \cite{grassetti2020}, who applied Bayesian approaches to improve RAPM by incorporating prior information about players, leading to more stable estimates in the presence of sparse or noisy data. 
Moreover, they have introduced the estimation of lineup effects as an additional factor (positive or negative) for each player, depending on the lineup in which they are included.


Regarding the football (soccer) players' evaluation, a broad variety of research works exists. 
In the field of football player performance evaluation,  the contribution of Professor Hvattum is notable. 
He initially published a plus-minus methodology in football \citep{hvattum_pm_2015} --- adjusted plus-minus ratings obtained by a ridge multiple linear regression model --- which has been extended by suggesting an additional component in each player rating corresponding to the league they participate \citep{hvattum_pm_2019}.  
He also created and maintains a YouTube channel called ``Football Player Ratings''\footnote{\url{https://www.youtube.com/@footballplayerratings}}, where he presents and discusses statistical models for evaluating individual football players in a straightforward and accessible manner through clear, easy-to-understand videos.

In a later work, \cite{hvattum_2020} proposed an alternative plus-minus rating model, and introduced individual offensive and defensive contribution football players. 
By considering segments of a football match with fixed lineup, ratings are determined by minimizing the squared difference between the actual goals scored and the expected goals based on players’ ratings, accounting for effects like home-team advantage, segment duration, red card impact, age effect, and league strength. Consequently, the numerical ratings represent each player’s contribution to goals scored per 90 minutes. He, also, applied ridge regularization technique to obtain more stable ratings \citep{hvattum_2021}, while he proposed weights for each segment difference, depending on the timepoint of the segment, the duration of the segment, and the game states.

Despite the clear diversity of plus-minus ratings in performance evaluation, there remains a gap regarding the lack of use of logistic models in the field --- particularly within basketball analytics --- presenting an intriguing area for further investigation.
Consequently, based on the literature, the primary aim of this study is to develop multinomial logistic models for play-by-play data, inspired by the concept of Regularized Adjusted Plus-Minus (RAPM), by applying lasso as an alternative regularization technique and by considering a new weighted evaluation metric based on each player's participation in the game.

\section{Material and Methods}
\label{sec2_data_methods}

\subsection{Initial Raw Dataset}
\label{section_initial_data}
In this study, play-by-play possession NBA data were used from the 2021-2022 season. Data were collected from  NBA logs. 
The unit of study (i.e., observation) is each possession, for which points scored and the lineups of the two opposing teams are collected as features. 
The original (unprocessed) dataset contained 322,852 observations (possessions) and 13 variables. A description of the available variables follows:  
\begin{itemize}
\item[-] \textbf{home\_off:} A dichotomous variable that signifies whether the team in possession is the host or the visitor. 
\item[-] \textbf{pts:} This is the main outcome variable which records the points achieved in each possession taking values ranging from zero to six (0--6). Naturally, the occurrence of five or six points in one possession was very rare. Hence, it was reasonable that there were only two and one possessions with five and six points, respectively.
\item[-] \textbf{season\_type:} A dichotomous variable that shows the type of season the game occurred (regular season or playoffs).
\item[-] \textbf{O1--O5 and D1--D5:} 
String variables refer to the players (and their teams) participating in each possession. 
Each string consists of the team's abbreviated name and the player's name, e.g., LAC22 Ivica Zubac.
Features O1–O5 refer to the players of the team in possession (i.e., on offense), while D1–D5 refer to the players of the opposing team (i.e., on defense). 
\end{itemize}  

\subsection{Final Dataset}
\label{section_final_data}
The original raw dataset was appropriately transformed in order to be used in the subsequent analysis. 
Specifically, the players'-related variables (O1--O5 and D1--D5) were transformed into a series of dichotomous variables (one offensive and one defensive for each player who participated in at least one game of the NBA league).  
These dichotomous variables denote the presence or absence of each player in the field, in each possession, with the team in possession or the team in defence. 
For this dataset, we have considered a total of 717 players.  Thus, 
the final data matrix is an $n \times m$ sparse matrix with $n=322,852$ rows (possessions) and $m=1,437$ columns with the features used in the analysis. 
The total set of features consists of $717 \times 2$ binary variables used to evaluate the offensive and defensive contribution of each player (i.e. each player is split into their offensive and defensive coefficients) plus three extra columns from the original raw dataset: \texttt{home\_off} (binary), \texttt{season\_type} (factor with 2 levels), and \texttt{pts} (points per possession).

The number of points  scored by the team in possession (pts) is the main outcome (or response) variable of this study. Therefore, the dichotomous variables $X^o_j$ of each player $j$ in the team in possession are coded as zero/one (0/1), while the corresponding variables $X_j^d$ of player $j$ in the defending team are coded as zero/minus one (0/-1); for $j\in \{1,2,\dots,K\}$ with $K$ denoting the total number of players in our dataset. This coding scheme reflects the positive and negative (or opposite) effect of each player on the outcome variable.

Concerning the outcome variable (pts), no points were scored in most of the possessions ($59.6\%$). Of the remaining ones, $26.2\%$ resulted in two points and $11.6\%$ in three points. 
One-point possessions accounted only for $2.5\%$ of the total number of possessions. 
This was due to the fact that two successful free throws (in one possession) were considered as two-point possessions.
Finally, more than three points were scored in only $0.1\%$ of the possessions. 

The number of possessions is almost balanced between the home and away teams, with a slight advantage for the away team (278 more possessions out of 322,852 total possessions). 
This is reasonable and expected due to the sequential nature of the game. 
Regarding the type of the game, most of the possessions ($93.2\%$) occurred in the regular season, as the NBA league format requires each team to play 82 games in the regular season and only 4--28 games in the playoffs.

\subsection{Models and Analysis}
\label{section_models}
In order to construct a player evaluation metric using statistical models, we require an efficiency measure as a dependent variable. Usually, in sports, this variable is the final outcome or score. 
For our study, which uses play-by-play data, the efficiency factor (or dependent or outcome variable) corresponds to the number of points scored by the offensive team in each possession.

In this section, we start from the standard approach in the basketball bibliography for obtaining RAPM measures which is based on the normal regression model using ridge regression. The main drawback of such a model is that the normal distribution attached to it is incompatible with the use of the number of points scored in each possession as the dependent efficiency variable. Therefore, we proceed with models that are more suitable for this kind of response.  Initially, we use a binomial logistic regression model with the binary indicator of scoring or not scoring per possession as a dependent efficiency variable. 
Although this model is more appropriate mathematically than the normal regression model, there is a loss of information since all different types of scoring are merged into one unified category. This loss of information (i.e. the different types of scores) is quite important for describing the nature of basketball. For this reason, we proceed by proposing a multinomial regression logistic regression model which takes into consideration the three different outcomes of a possession.

In all the above models, additionally to the ridge approach, we have also implemented lasso which has the characteristic that places specific players in an ``average" category. Overall, lasso seems to outperform the ridge approach in many different perspectives that will be later explained in Section \ref{sec3_results}. 

Finally, considering our reviewers' suggestions, we have also included in our comparisons a Poisson regression and an ordinal logistic model. The results of these models confirmed our initial intuition that such models are not appropriate for the specific case under study.

\subsubsection{Low-time players}
\label{section_ltps}
Prior to proceeding with the implemented models, we need to introduce the concept of {\it Low-time players (LTPs)}. In the following, we use this term for the players who participate for less than a threshold of minutes per game, usually due to their restricted role in the team. Their performance may affect the evaluation metrics that are widely used to assess the contribution of basketball players to their teams' success. For instance, conventional evaluation metrics do not take into consideration the context and quality of their playing time. Moreover, such metrics heavily depend on the performance of the other players on the court, as well as the quality of the opposing team\footnote{\url{https://www.nbastuffer.com/analytics101/plus-minus/}}.
From this perspective, \cite{hvattum_2019} argues that the quality and the performance of the team that a low-time player participates greatly affects their plus-minus ratings.
As a result, poor players on good teams are overrated, while good players on poor teams are underrated. 

A common way to resolve the problem is to impose a threshold on the playing time and exclude the corresponding LTPs from the calculation of the performance evaluation metrics; see for example \cite{rosenbaum_2004} and \cite{ilardi_barzilai_2008}. 

Subsequently, in the following, we consider as LTPs all players with playing time less than 200 minutes for the whole of the season. 
This threshold,  selected to be between the recommended thresholds used by \cite{rosenbaum_2004} and \cite{ilardi_barzilai_2008},
identifies 232 (out of 717) as LTPs which is the $32\%$ of the players in our dataset.

\subsubsection{Linear normal regression with shrinkage methods}
\label{section_linear}
We first proceed with the standard approach which uses the normal regression model. 
Hence for considering $K=717$ different players, the model is written in the following way
\begin{equation}
    Pts_i \sim N( \mu_i, \sigma^2) \mbox{~~with~~}
    \mu_i = b_0 + \sum_{j=1}^p b_{j} X_{ij}+ \sum_{k=1}^K \beta^o_{k} X_{ik}^o +\sum_{k=1}^K \beta^d_{k} X_{ik}^d, 
\label{normal}
\end{equation}
for $i=1, \dots, n$; where $n$ is the total number of possessions in our dataset, 
$K$ is the total number of players under consideration, 
$Pts_{i}$ is the number of points achieved in possession $i$, 
$\mu_i$ is the expected points for possession $i$ for this model, 
$X_{ik}^o$ and $X_{ik}^d$ are the indicators of player $k$ in possession $i$ for the offensive and the defensive team, 
$X_{ij}$ refers to the extra covariate $j$ for possession $i$, $p$ is the  number of extra covariates used in the model formulation, 
$\beta^o_{k}$ and $\beta^d_{k}$ are the offensive and defensive, respectively, RAPM coefficients for player $k$, 
$b_{k}$ are the effects of any extra covariates and 
$b_0$ is the overall constant of the model (usually referring to a reference average player obtained by players with zero or very close to zero RAPM coefficients). 
Finally,  $ \sigma^2$ is the error (or unexplained) variance which specifies the accuracy of the fitted model. 
In many published articles \citep[see for example][]{ilardi_barzilai_2007, ilardi_barzilai_2008}, the number of points per possession is multiplied by a factor of 100.

The additional covariates in the above formulation can be any data that enhance the forecasting of the final outcome of each possession. 
In our study, we used the following extra covariates: the home advantage of the team, the game type (regular season or playoff), and the teams involved in each game and possession.
Nevertheless, none of these covariates were found to have a significant effect on the points in addition to the player's RAPM coefficients. This result is consistent with the bibliography where they primarily focus on the estimation of RAPMs without introducing any additional information to the model; see, for example, in \cite{ilardi_barzilai_2007, pelechrinis_2019}. 

Hence interest lies in the estimation of $\beta_k^o$ and $\beta_k^d$ coefficients (i.e. on ORAPM and DRAPM, respectively). 
The constant term $b_0$ here is a nuisance parameter with no direct or meaningful interpretation under the current formulation. 
The  RAPM coefficients $\beta^o_k$ and  $\beta^d_{k}$ measure the offensive and defensive, respectively, contribution of player $k$ for $k=1, \ldots, K$ players. 
Pairwise comparison between two players (denoted by $k$ and $m$) can be achieved by considering the RAPM differences 
$\beta^o_k - \beta^o_m$, $\forall k,m=1, \ldots, K, k \neq m$.
These differences represent the mean discrepancy in the expected points scored by each player's team when player $k$ substitutes player $m$ in a specific possession, with all other players on the court staying the same. 
Similarly, the difference between two defensive RAPMs $\beta^d_k - \beta^d_m$ ($\forall k,m=1, \ldots, K, k \neq m$) expresses the average discrepancy in points of each player's team conceded (by the opponent team) when players $m$ play instead of player $k$ with all other players on the court staying the same. Hence, the larger the difference is, the higher the offensive and defensive contribution of $k$ player.

\subsubsection*{Ridge Regression} 

\vspace{-0.5em}
The related literature mostly suggests ridge regression to derive RAPM ratings. Ridge regression shrinks coefficients towards zero but does not set them exactly equal to zero. This has the advantage that every player will get a non-zero evaluation coefficient for his offensive and defensive contributions; 
for more details about the ridge regression see \cite{ridge} and in \cite{ilardi_barzilai_2007, hvattum_2019, plos_rapm2020} for the implementation in obtaining basketball RAPMs evaluation metrics. 

In the implementation of ridge regression, for the estimation of RAPM coefficients, we have used 10-fold cross-validation in  order to specify the tuning shrinkage parameter of ridge regression (often denoted by $\lambda$). 
In the ridge bibliography, one of the standard choices is to use  $\lambda$ which minimizes the average mean root square error (RMSE) obtained from cross-validation. This value was found to be equal to $\lambda_{min}=241.07$ which is considerably lower than the standard value of 2000 reported in RAPM bibliography \citep[see for example][]{sill_2010}. 
The shrinkage percentage with the choice of $\lambda_{min}$ was extremely high ($>99\%$). 

A clear issue that emerged in the RAMP-based player rankings,
is that low-time players (LTPs), namely players with less than 200 minutes played in the entire season, appear at the top of the evaluation list with exceedingly high RAPM ratings.
This is in line with the literature; refer, for instance, to in \cite{oliver_2002}.
To address this issue, we opted to implement lasso instead of ridge and, in a second step, to exclude LTPs from the analysis or to eliminate their RAPMs from the regression evaluation metric.

\subsubsection*{Lasso Implementation} 

\vspace{-0.5em}
Lasso has the extra property, in comparison to ridge, that shrinks smaller coefficients exactly to zero. 
This provides a more meaningful interpretation of the intercept coefficient of the implemented model, since now it represents the average contribution  of a reference player in each possession. 
By reference player, we mean any player belonging to the group of players with zero coefficients.
A total of 553 offensive and 574 defensive players (out of the 717) have been identified to belong in this reference group. 
Therefore 184 offensive and 143 defensive players were found with non-zero lasso RAPM ratings.  
To derive the lasso RAPM ratings we have used  the shrinkage parameter value of $\lambda_{min}=0.27$  obtained by a 10-fold cross-validation.

While there is a noticeable improvement in the lasso RAPMs when compared to the ones obtained by ridge, low-time players are still present in the top performances. 
Specifically, a smaller fraction of LTPs (as defined in this section) are among the top 100 RAPM performers when using lasso rather than ridge, accounting for 14\% and 77\%, respectively. 
Regarding the top 20 offensive and defensive players, the ridge ratings assign all positions to LTPs, whereas for the lasso ratings, only $20\%$ of the top-20 include  LTPs (all of them in the first five positions).

We now proceed by examining the top 50 lasso RAPM players. 
From this analysis, we reach the following  conclusions:
\begin{itemize}
    \item[--]For the offensive RAPMs, 
    out of the 28 top-voted All-Star players (for 2021-22), 16 were found in the top-50 and 13 in the top-28 of lasso RAPM ratings whereas for ridge only four and none, respectively.
    Note that, the selection of 28 players is based on the large voting-score gap between the $28^{th}$ and the $29^{th}$ player (11.5, while the subsequent player in the ranking has a score of about 40).

    \item[--]
    For the defensive RAPMs,  
    by taking into account the two ``All-defensive'' teams (two rosters of five players each) of the season 2021-2022, five out of 10 players are in the lasso RAPM top 50 while the remaining five players belong to the reference group. On the other hand, the ridge RAPM top 50 does not include anyone of these players.
\end{itemize}

\subsubsection{Filtering low-time players (LTPs)} 
\label{section_Filtering_ltps}
\vspace{-0.5em}
 Although lasso outperforms ridge,  some LTPs still seem to have an impact on the evaluation metrics. 
 A possible solution to improve the lasso ratings might be to exclude LTPs from our analysis as already suggested in the relevant bibliography.
Therefore, we have implemented both regularization methods (lasso and ridge) after
 removing 232 players with less than 200 minutes played (filtered dataset). 
This threshold value is comparable with similar choices in the literature: \cite{rosenbaum_2004} used data for players with more than 250 minutes played in two seasons 2002-2004, while \cite{ilardi_barzilai_2008} used a higher threshold of more than 300 minutes played in the 2007-2008 season. 

Before removing LTPs from the dataset, we implemented an intermediate step where all players with a playing time of less than 200 minutes (LTPs) were considered as a reference group with the same RAPM. The ratings obtained from this intermediate step were found to be similar and highly correlated with those obtained after their removal. Therefore, we decided to proceed with the latter analysis, which is commonly adopted in the literature. 

The filtered dataset now consists of 485 players (out of 717 in the original dataset) in which we obtain a total of 970 RAPMs in total (offensive and defensive ones). 
In this filtered dataset, estimated ridge and lasso RAPMs exhibit higher correlation ($78.8\%$ and $73.9\%$ for offensive and defensive RAPMs, respectively) than those obtained from the full dataset  ($60\%$ and $55.2
\%$, respectively). This correlation further increases to $90\%$ if we consider only the non-zero lasso RAPMs. 

Although results improved in both approaches when using the filtered dataset, the ridge-based RAPMs did not show sufficient improvement, as they tend to prioritize unexpectedly performed players at the top of their rankings. For instance, only $28\%$ of the top 100 offensive and defensive ridge RAPMs consisted of starter players. This marks a substantial improvement compared to the corresponding percentage of $8\%$ when all players were included in the analysis. 
However, this improvement does not appear to be adequate.

\begin{figure}
    \centering
    \includegraphics[width=0.6\linewidth]{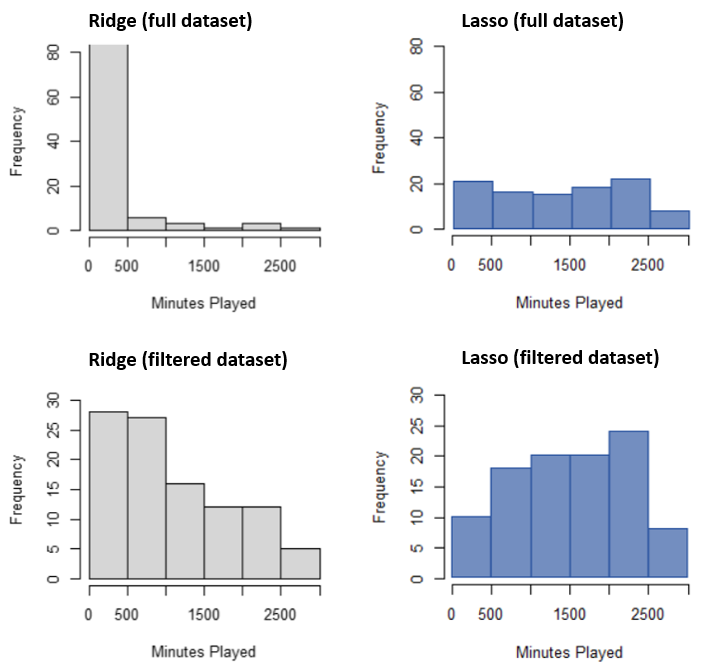}
    \caption{Minutes played of top 50 offensive and 50 defensive ridge (grey) and lasso (blue) RAPM players for the full (first row) and the filtered dataset (second row).}
    \label{fig:MP_lm_dataset}
\end{figure}

On the other hand, lasso-based RAPMs showed a greater improvement. For instance,  $42\%$  of the top-50 offensive and defensive RAPMs were starters in the full dataset, while for the filtered dataset this percentage increased to $55\%$. Furthermore, the distribution of the playing time for the top-50 players in the filtered dataset was left-skewed. This implies that the majority of the top-rated players had high playing time (Figure \ref{fig:MP_lm_dataset}).

Despite the improved promising results of the lasso RAPMs in the filtered dataset,  we will further explore the implementation and the development of RAPMs based on the multinomial logistic regression model which is more appropriate for the outcome variable of interest here i.e. the number of points scored per possession.  In the following, we continue our analysis using the filtered dataset.
Before we proceed to the more complex case of the multinomial logistic regression model, we will first implement the simpler binary logistic model, where the response variable is simply whether the team in possession scored or not.
This is the first step towards the final goal of fitting the more advanced multinomial logistic regression model. As we will show later, this also allows us to provide an indirect approximate interpretation for the simple regression-based approach.

\subsubsection{Logistic regression with shrinkage methods}
\label{section_logistic}
In this section, we proceed by considering the response variable $Y_i$, which records whether the team in possession scored or not. Since this simplified response is now binary, we will consider a standard, Bernoulli-based, logistic regression model given by 
\begin{eqnarray}
Y_i &=& {\cal I}(Pts_i >0) , \mbox{~~with ~~}  
Y_i \sim Bernoulli (P_{i}) \mbox{~~ and ~~} P_{i}=\frac{1}{1+e^{-\mu_i}}
\label{binomial}
\end{eqnarray}
for $i=1, \dots, n$;  where  $\mu_i$ is the linear predictor as defined in the normal linear model formulation \eqref{normal}, ${\cal I}(A)$ is the indicator function taking the value of one if $A$ is true and zero otherwise and  $P_{i}$ is the probability of scoring in $i$ possession.

The estimated coefficients $\beta_k^o$  and $\beta_k^d$of each $k$ player's offensive and defensive dummy variables $X_{ik}^o$ and $X_{ik}^d$ are measures of their contribution to their team’s scoring ability (in the form of binary outcome). These coefficients can be considered as RAPMs on a different scale since they measure the effect of each player in terms of log-odds of scoring. 

For the estimation of logistic regression RAPMs, we have similarly applied regularization methods as in the normal-based RAPMs. 
For both ridge and lasso methods, the RAPM-based ranking where obtained using the shrinkage parameter value $\lambda_{min}$  obtained from a 10-fold cross-validation. 
The lasso implementation resulted in 617 player RAPMs (either offensive or defensive) which were shrunk to zero. 
This represented approximately the $63.8\%$ of the total estimated RAPMs (and players, respectively).

From binomial analysis, the correlation between RAPM ratings derived from the ridge binomial (\(\text{RAPM}^\text{B}\)) and ridge Normal (\(\text{RAPM}^\text{N}\)) methodologies reveals an interesting result. 
Specifically, from our analysis, we found a significant linear relationship between these two RAPM ratings
with $R^2\approx 0.89$. 
This strong linear association between the two approaches provides a reasonable theoretical justification for using the normal RAPMs as approximations of the binomial RAPMs, which are derived from a model with a properly specified distribution.
Moreover, the estimate equation\footnote{$\text{RAPM}^\text{B}_k = 0.02 \times \text{RAPM}^\text{N}_k + \varepsilon_k, ~  \varepsilon_k \sim \mathcal{N}(0, 0.008^2)$} allows us to compute the binomial RAPMs from those derived under the normal distribution. 



Although the logistic regression RAPMs can be considered an improvement when compared to the normal-based RAPMs, it does not consider the scoring information that is inherent in basketball, as they disregard the exact number of points scored. 
Therefore, this model should be further extended to account for the different number of points scored in each offence/possession. 
This is crucial, as each scoring category (one, two or three-pointers) demands different skills from the players and strategies from the team.

\subsubsection{Multinomial logistic regression with shrinkage methods}
\label{section_multinomial}

In this section, we proceed with the implementation of a multinomial logistic regression for modelling the scoring outcome of each possession. As we have already discussed, this approach is more appropriate for our problem since it can model separately each of the types of scores. In this way it accounts for the different scoring contributions of each player. 

Due to the large size of the data, for computational efficiency, the multinomial model was fitted indirectly by using three binomial models for the three scoring categories under consideration (one, two and three or more points per possession versus no points).

Following the methodology implemented in normal models, for each of the three models, lasso regularization is used as a shrinkage method with the penalty parameter $\lambda$ set to the value which attains the minimum RMSE ($\lambda=\lambda_{min}$)\footnote{The choice of $\lambda_{1se}$, resulted in a degenerate model in which all player RAPMs were shrunk to zero. Consequently, this model is practically useless for evaluating and ranking players.}. 

Hence, we consider the following model formulation 
\begin{equation*}
Y_i^{\mathcal M} = Pts_i, \mbox{~for~} Pts_i \in \{ 0, 1, 2\} \mbox{~and~} Y_i^{\mathcal M} = 3, \mbox{~for~} Pts_i \ge 3    
\end{equation*}
with
\begin{equation}
Y_i^{\mathcal M} \sim Multinomial ( \boldsymbol{\pi}_i ) \mbox{~and~}
\boldsymbol{\pi}_i = ( \pi_{i0}, \pi_{i1}, \pi_{i2}, \pi_{i3}) \mbox{~with~} \sum_{\ell=0}^3 \pi_{i\ell}=1. 
\label{multinomial}
\end{equation}
The multinomial probabilities are given via the following equations
\begin{equation}
\pi_{i\ell}=\frac{e^{\mu_{i\ell} }}{1+e^{\mu_{i1} }+e^{\mu_{i2} }+e^{\mu_{i3} }}  \mbox{~for~} \ell=0,1,2,3; 
\label{probSystem}
 \end{equation}
where $\pi_{i0}, \pi_{i2},\pi_{i2}, \pi_{i3}$ are the probabilities of no scoring, scoring one point, two points or three points (or more) in $i$ possession. 
The linear predictor $\mu_{i\ell}$ is given by 
\begin{equation}
     \mu_{i\ell}= b_{0\ell} + \sum_{j=1}^p b_{j\ell} X_{ij}+ \sum_{k=1}^K \beta^o_{k\ell} X_{ik}^o +\sum_{k=1}^K \beta^d_{k\ell} X_{ik}^d, \mbox{~for~} \ell =1,2,3 \mbox{~and~} \mu_{ik}=0, 
\label{eqSystem}
\end{equation}
for $i=1, \dots, n$. 

We implemented the multinomial regression formulation by considering three separate binomials given by 
\small 
\begin{eqnarray}
Y_i^{(\ell0)} \!\!\!  &=& \!\!\!  \left\{
\begin{array}{ll}
{\cal I}(Pts_i = \ell) &\!\! \mbox{when~} Pts_i \in \{0, \ell\} \\ 
NA                     &\!\! \mbox{otherwise}
\end{array}\right. \hspace{-0.5em}, \mbox{~with~}  
Y_i \sim Bernoulli \big(P_{i}^{\ell0}\big) \mbox{~and~} \log \frac{P_{i}^{\ell0}}{P_{i}^{00}}= \mu_{i\ell}. ~~~
\label{binom_system}
\end{eqnarray}
\normalsize 
From the above-fitted models, we obtain the multinomial probabilities using \eqref{probSystem}.  
Note that the estimates based on the separate logistic regression approach are less efficient with larger standard errors than the direct multinomial logistic approach. Nevertheless, the loss is minor when the baseline category is dominant on the data like in our case \cite[Chap. 8]{agresti}.
We prefer the separate logistic regression approach here for two reasons. 
First, the separate regression approach offers advantages in terms of computational efficiency. 
Hence, it can be implemented for datasets with large sample sizes, as in the present study.
Secondly, this approach allows us to implement different variable selection and shrinkage algorithms on the effects of each scoring category. 
In this way, we can identify the contribution of each player in the three different types of scores.

Team effects were not included in the fitted model, since they were not found to be significant. 
In the final multinomial model, 362 offensive and 316 defensive players actively contributed to the model. 
These were players with non-zero RAPMs in at least one of the binomial components. 
In this way, if a player has a significant impact in at least one scoring situation, his effect is taken into consideration in the final model formulation. 

Although our approach is beneficial and more informative than the simple regression or logistic regression approach in the sense that we identify the contribution of each player at the different types of scores, in the end, we would like to summarize the offensive and the defensive contribution of each player with an overall index. 
For the offensive contribution of a player, this can be achieved by calculating the expected number of points (in a single possession) scored by the team of this player when he is included in the lineup 
and all other players on the court are from the reference category (with zero RAPM) -- denoted by $EPTS^{o}_{k}$ . Similarly, the defensive contribution of a player can be evaluated by the expected number of points (per possession) conceded from the team of this player when he is not included in the lineup and all other players on the court are from the reference category (with zero RAPM). Equivalently, this is denoted by $EPTS^{d}_{k}$  
Hence, the EPTS ratings are defined as 
$$
EPTS^{r}_{k}=E(Pts_i | X_{ik}^r \neq 0, {\bf X}^r_{i\setminus k} = {\bf 0}, {\bf X}^{\overline{r}}_{i} = {\bf 0})  \mbox{~for~} r \in \{o, d\}. 
$$
for $k=1,\ldots,485$. 
The two measures of the expected number of points, 
$EPTS^{r}_{o}$ and $EPTS^{r}_{d}$,  will be functions on the offensive and defensive coefficients of the fitted multinomial regression model. 
Under this perspective, the expected number of points (EPTS) will be obtained by 
\begin{equation}
EPTS^{r}_{k}=\Pi^{r}_{k1}+2\times \Pi^{r}_{k2}+3.01\times \Pi^{r}_{k3}, ~r \in \{o, d\}; ~~\overline{r} =  \{o, d\} \setminus r 
\label{epts}
\end{equation}
with  $\Pi^{o}_{k\ell}$ for $\ell\in\{1,2,3\}$ being estimated by the multinomial logistic regression probabilities of scoring one, two or three points and more, respectively\footnote{The value of 3.01 (instead of 3) is simple the average points for all possessions with points more or equal to three.}, per possession from the team of player $k$ when he is included in the lineup under a specific simplified scenario. 
This scenario is the case where all teammates of player $k$ in the lineup and opponents are players included in the reference group (i.e. with zero regression coefficients). 
Similarly, $\Pi^{d}_{k\ell}$ (for $\ell\in\{1,2,3\}$) are the probabilities of scoring one, two or three points and more, respectively, per possession by the opponent of the team of player $k$ when he is not included in the lineup under the same simplified scenario. 
Under this simplified scenario, the point probabilities are given by  
\begin{equation}
\Pi^{r}_{k\ell}=\dfrac{e^{\mu^{r}_{k\ell} }}{1+e^{\mu^{r}_{k1} }+e^{\mu^{r}_{k2} }+e^{\mu^{r}_{k3} }}  
\mbox{~~with~~} 
\mu^{r}_{k\ell}= b_{0\ell} + b^{r}_{k\ell} \, , 
\label{playerMU}
\end{equation}
for $\ell \in \{ 1,2,3 \}$. 
The above overall evaluation score is now based on the RAPM coefficients of the specific type of points and additionally to the constant terms which adjust calculations over the simplified scenario where the rest of the players come from the reference group (with zero RAPM coefficients). 

For instance, when comparing two players, A and B, we consider the following statements:
\begin{itemize}
    \item[--] In the scenario where \( EPTS^o_A = 0.9 \) and \( EPTS^o_B = 1 \), Player B contributes one (1) point per possession to his team's offensive performance when he is on the court, which is higher than Player A's contribution of 0.9 points per possession, considering that all other players on the court are the same. This indicates that Player B has a greater offensive impact, as his presence in the lineup results in a more favorable scoring outcome for his team compared to the corresponding scoring outcome for Player A.
 
    \item[--] Regarding the defensive EPTS measures, let us consider the case where \( EPTS^d_A = 0.9 \) and \( EPTS^d_B = 1 \). 
    Then, Player B's defensive contribution is measured as one point conceded per possession when he is not playing. This is compared to 0.9 points conceded per possession when Player A is not participating in the specific possession, assuming that all other players on the court remain the same.  
    This indicates that Player B's absence has a greater impact on the team’s defense, as the team concedes a higher number of points per possession when he is not on the court. Consequently, this suggests that Player B is more valuable to his team's defense than Player A.

\end{itemize}

By defining EPTS as described above, player comparisons in both offensive and defensive aspects of the game become straightforward and easy to interpret, by using a single metric for each case rather than three separate ones for the probabilities of scoring or conceding one, two, three, or more points. 
Moreover, the interpretation of offensive and defensive ratings is consistent: a higher EPTS score always signifies a greater contribution in both scenarios. Therefore, the higher the EPTS value, the greater the player’s (offensive or defensive) impact on his team.

\begin{figure}
    \centering        
    \includegraphics[width=0.6\linewidth]{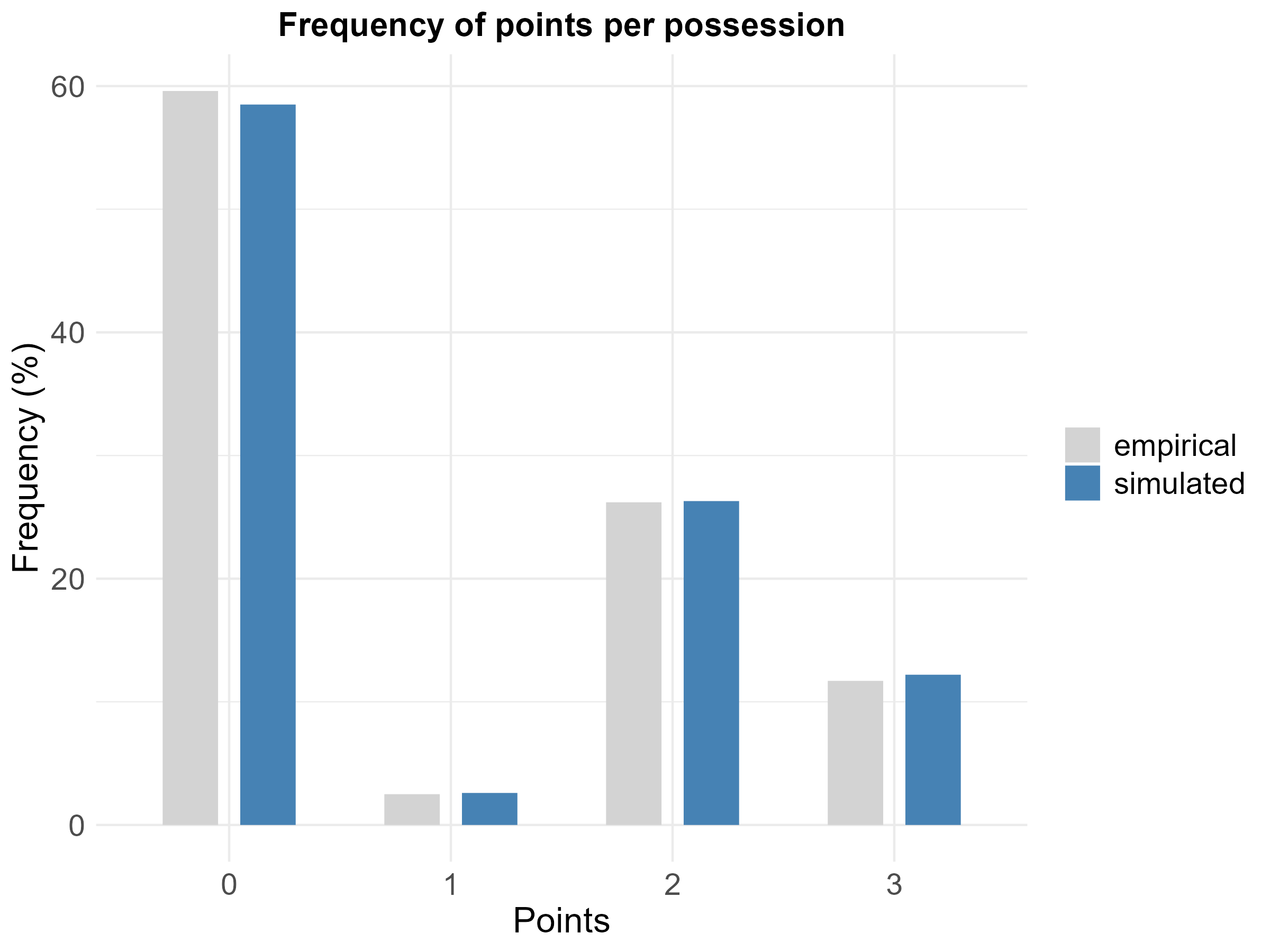}
    \caption{Relative frequencies of actual points per possession vs. the fitted ones estimated by the lasso multinomial model using simulation.}
    \label{fig:Multinomial_fit}
\end{figure}

The Root Mean Squared Error (RMSE) of the fitted multinomial model is lower compared to the normal model ($1.62<1.86$). 
To further assess the model's fit, we simulated 1,000 samples from a multinomial distribution using the estimated probabilities per possession. The resulting relative frequencies of points per possession closely resemble those observed in the actual data (Figure \ref{fig:Multinomial_fit}) at the marginal level.  
A bootstrap chi-square test revealed no statistically significant difference between the marginal fitted frequencies and the observed ones (p-value = $0.55 > 0.05$). This finding suggests a good fit of the model.

\section{Results and Comparisons}
\label{sec3_results}

This section presents the key findings obtained by applying the methodological procedures outlined earlier in this work. Our focus lies in comparing the RAPM  approaches presented in previous sections. We demonstrate the superiority of the multinomial lasso-RAPMs (EPTS and WEPTS) compared to the ones obtained with ridge and lasso normal models.
Prior to this comparison, we establish the external validation criteria that serve as the basis for evaluating the performance of the different approaches.

\subsection{External Validation Criteria}
\label{ext_val_criteria}

To evaluate the ratings generated by the implemented models, we compare them by using a series of validation criteria that can be thought as external and ``objective'', due to their non-involvement in the model formulation. 
While the offensive RAPMs yielded promising results, the performance of the defensive RAPMs offers room for improvement.
This might be due to the nature of defensive play in basketball. Unlike offensive performance, which can be more readily influenced by individual player skill, defensive success is often based on effective team cooperation and collaboration. 

To evaluate the performance of the different RAPM approaches, we select the following  set of external validation criteria: 
\begin{itemize}
    \item[1.] \textbf{All NBA Teams Criterion:} This criterion examines the percentage of top RAPM players per position who are included in the ``best 3 lineups'' (15 players in total) which are announced according to the official NBA website (\href{https://www.nba.com/}{https://www.nba.com/}). As top-ranked RAPM players, we consider the six best players for each position except for the centers, where only the top three RAPM-ranked players are included.  
    \item[2.] \textbf{Low Time Players Criterion:} We identify the bottom 50 players in terms of playing time and measure the percentage of top-50 RAPM players within this group. This criterion assesses whether the RAPM approach effectively reduces the influence of players with minimal playing time.
    \item[3.] \textbf{Starters Criterion:}  We consider the percentage of top-50 offensive and defensive RAPM-based metrics (total of 100 players) that are classified as ``starters''. All top-six most played players per team are considered as starters.
    \item[4.] \textbf{Top-50 Box Score Statistics Criterion:} This criterion identifies the top-50 players based on a variety of box score statistics (points, assists, and offensive rebounds for offence; defensive rebounds, steals, and blocks for defence). We then measure the percentage of these players who are included in the top-50 RAPM rankings.

\end{itemize}

\subsection{Lasso vs. Ridge RAPMs in Normal Models}
\label{ridge_vs_lasso}

As a first step in our analysis, we compare the lasso RAPMs with more traditional ridge-based RAPMs commonly implemented in basketball analytics literature. 
The findings are quite promising in favour of  the lasso approach. 
Specifically, the lasso method appears to outperform ridge regression based on the evaluation criteria established  in Section \ref{ext_val_criteria}.

An examination of Tables \ref{tab:normal_criteria} and \ref{tab:normal_boxscore_criteria} reveals the  superiority of lasso over ridge.
Specifically, focusing on the playing time of top-rated players (Criterion 2), as shown in Table \ref{tab:normal_criteria}, our analysis highlights a clear advantage of lasso RAPMs over ridge ones. 
We observe that 14\% of the top 100 players in the ridge RAPM rankings are among those with the lowest recorded playing time. This percentage drops to 3\% for lasso normal and 6\% for the OLS ratings after removing the zero-contributed players indicated by lasso. 
The latter will be referred to as after-lasso RAPMs. 
This suggests a more accurate performance of the lasso models since it is generally unexpected for players with minimal in-season playtime to appear in the list with the highest contributions. This is a common behaviour of the plus-minus ratings found in ridge RAPMs, but restricted in the lasso approach.

On the other hand, a strong presence of starters among top performers is highly desirable.
Examining the third criterion (Criterion 3a in Table  \ref{tab:normal_criteria}), we observe that 55\% and 50\% of the players with the highest contributions are starters according to the lasso and after-lasso RAPMs, respectively. These percentages are about double (increased by 96\% and 79\%, respectively) the corresponding percentage  (28\%) achieved when using  the ridge RAPMs.  

While a high proportion of starters among top performers is desirable, the presence of starters among low performers requires further examination (Criterion 3b in Table \ref{tab:normal_criteria}). 
Interestingly, the lasso RAPM ratings include a higher percentage of starters (23\% and 19\% for lasso and after-lasso, respectively) compared to ridge RAPM (7\%). 
However,  this observation does not necessarily favour the ridge approach.
A possible explanation is that, on some occasions, a player might be included in the starting roster due to a lack of better alternatives or his role in the team might be offensive and not defensive (or vice-versa). 

\begin{table}[h!]
    \caption{Comparison of validation Criteria 1--3 between RAPM ratings for the Normal models.}
    \label{tab:normal_criteria}
    \centering
    \begin{tabular}{|>{\raggedright\arraybackslash}p{3cm}|c c c c|}
    \hline
   \textbf{Model} & Criterion \#1 & Criterion \#2 & Criterion \#3a & Criterion \#3b\\ [0.5ex]
    \hline\hline
    Ridge Normal & 27\% & 14\% & 28\% & 7\%\\
    Lasso Normal & \textbf{40\%} & \textbf{3\%} & \textbf{55\%} & 23\%\\
    Normal (after-lasso) & 33\% & 6\% & 50\% & 19\%\\
    \hline
    \multicolumn{5}{p{11cm}}{\footnotesize \it Criterion \#1: All-NBA teams, Criterion \#2: Low-time players in 100 top-RAPM, Criterion \#3a: Starters in 100 top-RAPM and Criterion \#3b: Starters in 100 bottom-RAPM; {\bf Bold} indicates the maximum value by column/criterion}
    \end{tabular}
\end{table}

An examination of the box-score statistics of the top-rated RAMP players (Criterion 4 in Table \ref{tab:normal_boxscore_criteria}) further strengthens the case in favor of  lasso RAPMs. 
From this table, we observe a clear advantage for lasso RAPMs, particularly when compared to the performance of ridge based top players  in key metrics such as points scored, assists, defensive rebounds, and blocks. 

\begin{table}[h!]
    \caption{Comparison of External Validation Criterion 4 for Lasso Normal ratings.}
        \label{tab:normal_boxscore_criteria}
    \centering
    \begin{tabular}{|>{\raggedright\arraybackslash}p{3cm}|c c c c c c|}
    \hline
   \textbf{Model} & PTS & AST & OREB & DREB & STL & BLK\\ [0.5ex]
    \hline\hline
    Ridge Normal & 30\% & 20\% & 12\% & 10\% & 18\% & 4\%\\
    Lasso Normal & 42\% & 26\% & \textbf{14\%} & 8\% & 14\% & \textbf{12\%}\\
    Normal (after-lasso) & \textbf{44\%} & \textbf{30\%} & 12\% & \textbf{20\%} & \textbf{22\%} & \textbf{12\%}\\
    \hline
        \multicolumn{7}{p{10cm}}{\footnotesize \it PTS: points scored per game, AST: assists per game, OREB: offensive rebounds per game, DREB: defensive rebounds per game, BLK: blocks per game, STL: steals per game; {\bf Bold} indicates the maximum value by column/box-score}
        \end{tabular}
\end{table}

\begin{figure}
    \centering
    \includegraphics[width=1\linewidth]{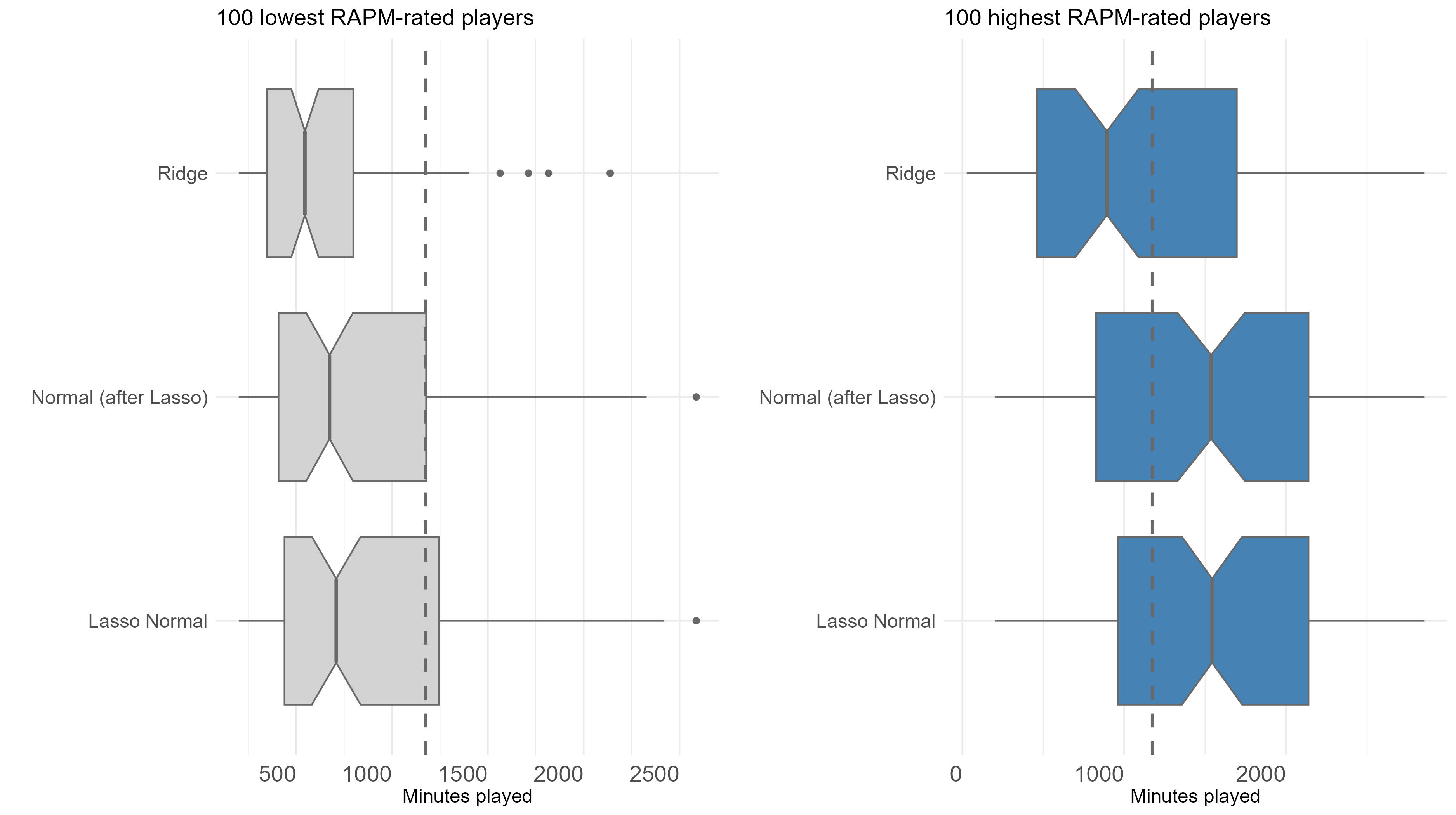}
    \footnotesize \it{Vertical dashed line is a threshold of the lowest playing time observed in starters.}
    \caption{Minutes Played of players in top and bottom 100 (50 offensive and 50 defensive) RAPMs.}
    \label{fig:normal_MP}
\end{figure}

\begin{figure}
    \centering
    \includegraphics[width=1\linewidth]{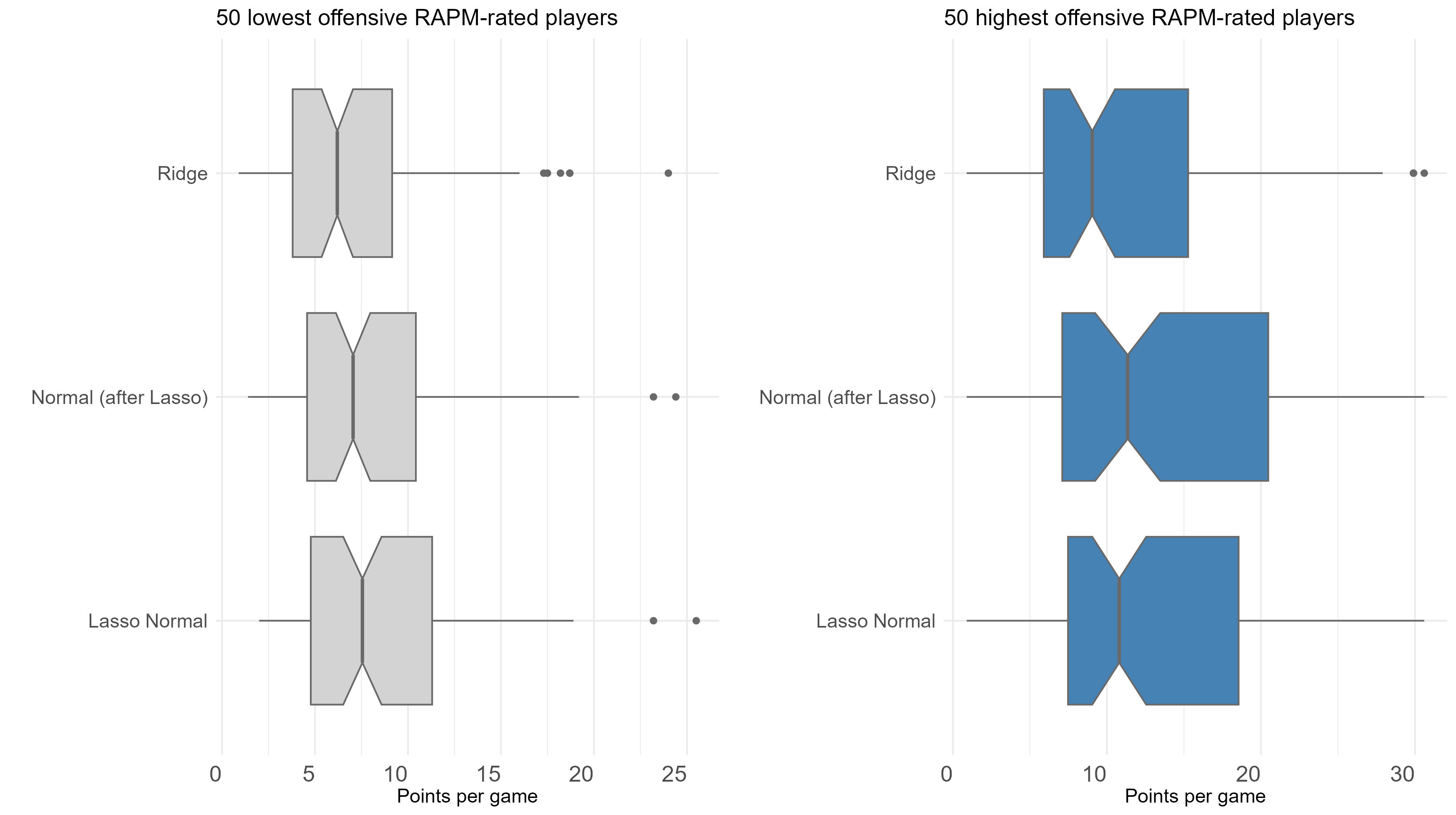}
    \caption{Points scored per game by players in top and bottom 100 (50 offensive and 50 defensive) RAPMs.}
    \label{fig:normal_PTSg}
\end{figure}

Further visual inspection of the distribution of the playing time (see Figure~\ref{fig:normal_MP})  suggests that the ridge model assigns high rankings to players with lower playing time, despite their lower scoring outputs (see Figure~\ref{fig:normal_PTSg}). This is also supported by the finding that  at least 75\% of the top 50 ridge RAPM players score no more than 15 points per game (Figure~\ref{fig:normal_PTSg}).
In contrast, lasso models prioritize players with both higher playing time and higher scoring performances. This is evident by the fact at least 75\% of the top performers in the offensive lasso (and after-lasso)  RAPMs score around 20 and 18 points per game, respectively.

To conclude with, based on the external validation criteria, our findings suggest that the lasso methodology offers ratings that lead to better player discrimination. 
Moreover, it yields more reasonable ratings compared to ridge regression. 
One potential drawback of ridge regression appears to be its handling of players with limited playing time. 
Ridge regression may overestimate or underestimate the performance of low-time players, potentially assigning them to extreme ranking positions (highest or lowest).
Based on the validation criteria, a comparison between the lasso RAPMs and the after-lasso ratings revealed no substantial differences. Nevertheless, in the following, we favour the use of the ratings obtained from the lasso approach. This preference is because the latter ratings 
are directly obtained while the free-of-bias (due to the penalty of the lasso method) estimates, after-lasso RAPMs, need extra computational effort after the initial screening procedure using lasso.

\subsection{Lasso Multinomial vs. Benchmarks }
\label{section_benchmarks_vs_multinomial}
After demonstrating the superiority of the lasso technique over ridge regression in Section \ref{ridge_vs_lasso}, we will now examine the proposed multinomial model, compared with the RAPMs obtained by two  ``benchmark'' models: 
(a) the lasso-normal and (b) the lasso-logistic. 

As discussed in Section \ref{section_multinomial}, the proposed multinomial model accounts for the discrete nature of points scored per possession. This choice, in combination with lasso and its shrinkage properties, offers a clear advantage over the commonly used ridge regression RAPMs. 
Therefore, in this section, we will examine and compare the external validation criteria (see Section \ref{ext_val_criteria}) for the the three models under consideration. 

From Tables \ref{tab:normal_multinomial_criteria} and  \ref{tab:normal_multinomial_boxscore_criteria}, 
we observe small differences in the performance, with no clear winner between the opposite distributions for the different evaluation criteria. 
The lasso multinomial performs better in two out of four evaluation criteria presented in Table \ref{tab:normal_multinomial_criteria}. 
When analyzing players from ``All NBA teams'' (Criterion 1), lasso multinomial outperforms both benchmark models by identifying a larger proportion of such players; 
$33\% < 40\% < 47\%$ for binomial, normal and multinomial, respectively. 
Furthermore, in Criterion 3b, the multinomial lasso performs marginally better than the rest cases, since fewer starters are included in the bottom 100 list ($20\% < 23\% < 44\%$). 
On the other hand, normal-based RAPMs outperform in Criterion 2, since a lower proportion of low-time players ($3\%$) is included in its top 100 player rankings. 
Finally, regarding Criterion 3a, a higher proportion of starters are included in the top-100 normal RAPM list ($55\%$). Regarding the box-score statistics (Criterion 4) of top-rated players, lasso multinomial is marginally better than normal lasso in all statistics except the number of points (PTS). 
In terms of differences in box score statistics between the multinomial and logistic regression models, the proposed model outperforms the binomial one in two criteria: offensive rebounds and steals, while the binomial model is better only in defensive rebounds; see Table \ref{tab:normal_multinomial_boxscore_criteria}.

\begin{table}[h!]
    \caption{Comparison of validation Criteria 1--3 between ratings for the lasso Normal, Binomial and Multinomial models.}
    \label{tab:normal_multinomial_criteria}
    \centering
    \begin{tabular}{|>{\raggedright\arraybackslash}p{2.5cm}|c c c c|}
    \hline
    \textbf{Model} & Criterion \#1 & Criterion \#2 & Criterion \#3a & Criterion \#3b\\ [0.5ex]
    \hline\hline
    Lasso Normal & 40\% & \textbf{3\%} & \textbf{55\%} & 23\%\\
    Lasso Binomial & 33\% & 6\% & 45\% & 44\%\\
    Lasso Multinomial & \textbf{47\%} & 9\% & 47\% & 20\%\\
    \hline
    \multicolumn{5}{p{11cm}}{\footnotesize \it Criterion \#1: All-NBA teams, Criterion \#2: Low-time players in 100 top-RAPM, Criterion \#3a: Starters in 100 top-RAPM and Criterion \#3b: Starters in 100 bottom-RAPM; \textbf{Bold} indicates the maximum value by column/Criterion} \\
    \end{tabular}
\end{table}

\begin{table}[h!]
    \caption{Comparison of External Validation Criterion 4 for lasso ratings of Normal, Binomial and Multinomial.}
    \label{tab:normal_multinomial_boxscore_criteria}
    \centering
    \begin{tabular}{|>{\raggedright\arraybackslash}p{2.5cm}|c c c c c c|}
    \hline
    \textbf{Model} & PTS & AST & OREB & DREB & STL & BLK\\ [0.5ex]
    \hline\hline
    Lasso Normal & \textbf{42\%} & 26\% & 14\% & 8\% & 14\% & 12\%\\
    Lasso Binomial & 38\% & \textbf{28\%} & 14\% & \textbf{20\%} & 16\% & \textbf{14\%}\\
    Lasso Multinomial & 38\% & \textbf{28\%} & \textbf{16\%} & 14\% & \textbf{20\%} & \textbf{14\%}\\
    \hline
    \multicolumn{7}{p{9.5cm}}{\footnotesize \it PTS: points scored per game, AST: assists per game, OREB: offensive rebounds per game, DREB: defensive rebounds per game, BLK: blocks per game, STL: steals per game; \textbf{Bold} indicates the maximum value by column/box-score} \\
    \end{tabular}
\end{table}

Based on the previous comparisons, we conclude that the lasso multinomial model performs somewhat better than the standard normal and binomial models.
Furthermore, the multinomial model is more realistic than the standard normal approach since the response is properly considered as a discrete random variable, whereas the normal approach (incorrectly)  assumes that the number of points per possession is a continuous random variable. 
In the case of the binomial model, the different types of points scored are not taken into consideration, which ignores important information about the basketball game.
Finally, the choice of the multinomial model is further supported by the fact that the RAPMs in conventional lasso models are shrunk to zero for around  $70.0\%$, and  $63.8\%$ of the players in the normal, and binomial case, respectively. In contrast, the corresponding proportion for the lasso multinomial model is reduced to only $30\%$. Since the goal is to develop an evaluation metric, reducing the RAPM for the majority of players to zero implies that, for a large number of players,  no clear evaluation and discrimination from the ``average'' level will be provided.

\subsection{Joint vs. Separate estimation} 
\label{section_jointVSseparate}
As discussed by \cite{hosmer2013applied}, joint estimation in multinomial logistic regression is generally more efficient than the separate logistic regression approach 
because it takes into consideration the entire dataset across all outcome categories, leading to more precise and stable parameter estimates. This approach, however, leads to a less flexible model specification, as it assumes the same sets of covariates across all categories. 

On the other hand, separate estimation offers greater flexibility by allowing for different relationships between predictors and each outcome category. This flexibility, while advantageous in certain scenarios, can result in higher variance in the estimates due to the reduced amount of data used in each independent model. Nonetheless, as noted by \cite{agresti}, this loss of efficiency is lower when the baseline category is dominant within the data, as is the case in our analysis. In such scenarios, the impact of larger standard errors is negligible, making the separate logistic regression approach a realistic alternative to the multinomial model, particularly when  minimizing the computational burden is set as a priority.

\begin{figure}
    \centering
    \includegraphics[width=0.8\linewidth]{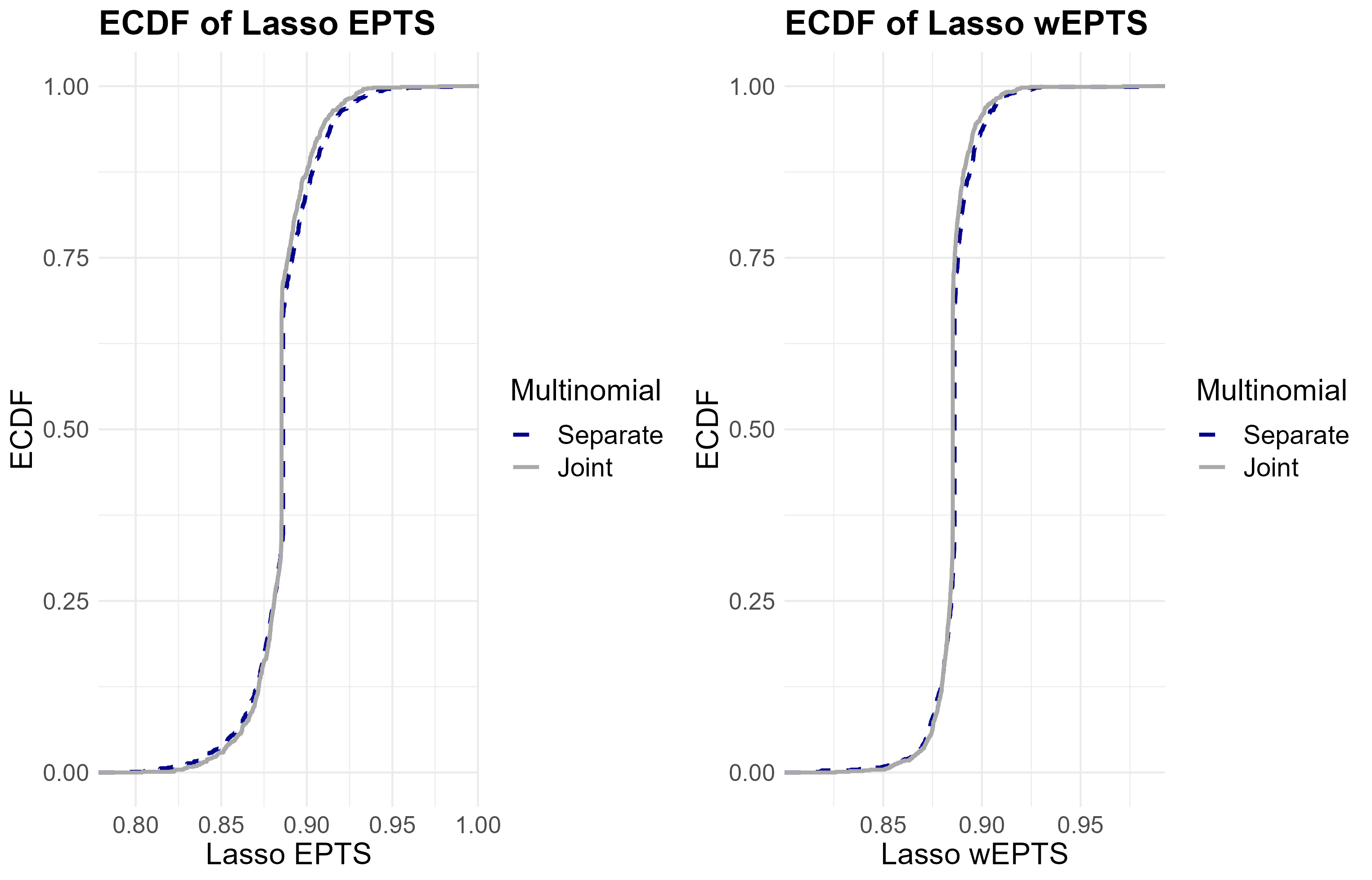}
    \caption{Empirical cumulative distribution function (ECDF) for the lasso EPTS and wEPTS obtained by the joint and separate estimation of the multinomial model.}
    \label{fig:ECDF_joint_vs_sep}
\end{figure}

\begin{figure}
    \centering
    \includegraphics[width=0.8\linewidth]{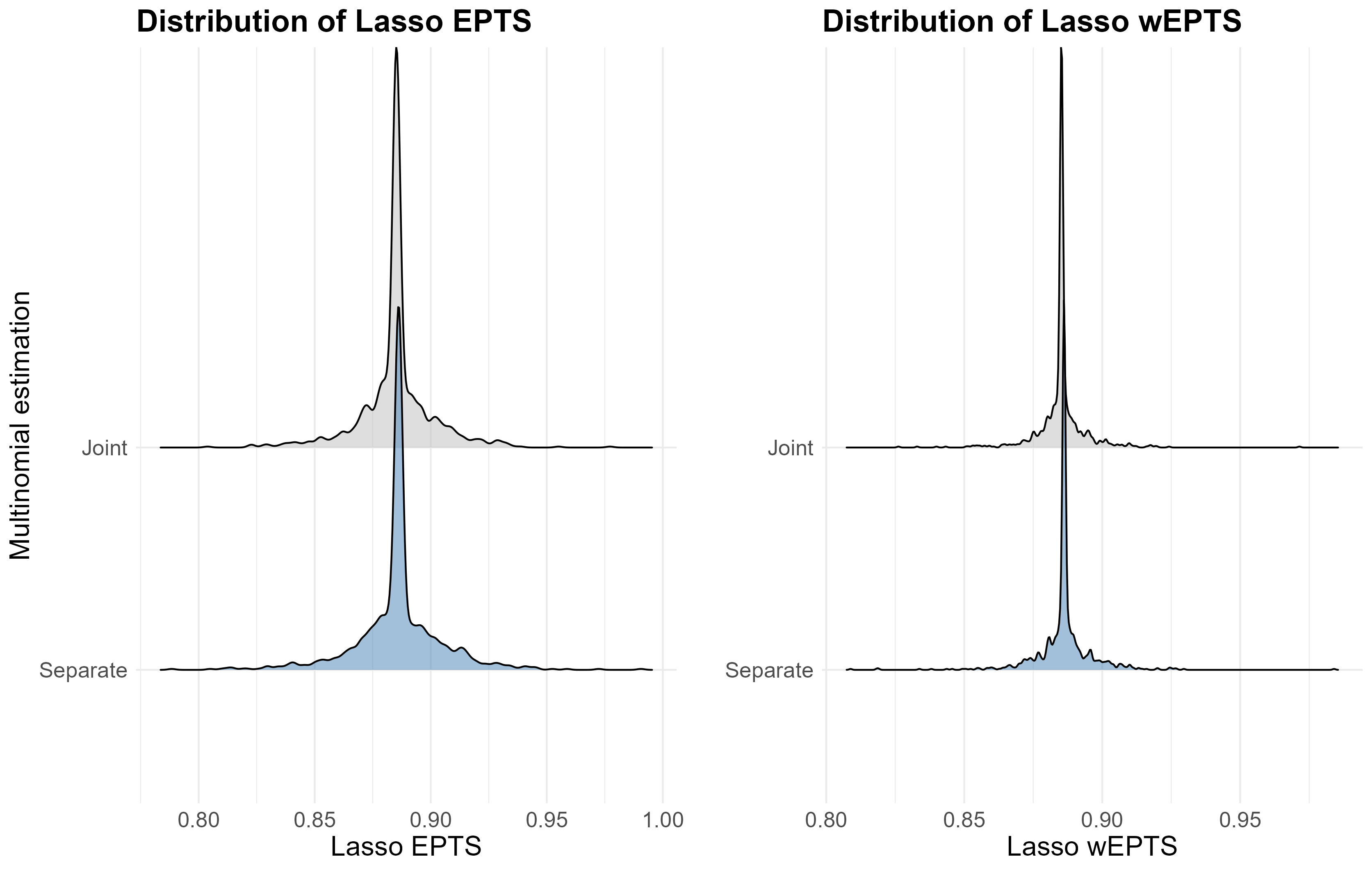}
    \caption{Distribution of the lasso EPTS and wEPTS obtained by the joint and separate estimation of the multinomial model.}
    \label{fig:distribution_joint_vs_sep}
\end{figure}

For the joint multinomial implementation,  a slightly higher number of players effect is shrunk to zero than the separate estimation (305 vs. 292; 31.4\% vs. 30\%). 
Moreover, the relationship between the Expected Points per Possession derived from the joint and separate approach is strong (Pearson's $r \approx 95\%$) while no significant differences are observed in the ECDF and distribution plots; see Figures \ref{fig:ECDF_joint_vs_sep} and \ref{fig:distribution_joint_vs_sep} respectively. 

Based on the results, it seems that the simplified EPTS ratings derived by separate multinomial regression approach offers an accurate and less demanding alternative to EPTS derived by the joint multinomial model approach.

\subsection{Poisson implementation}  
\label{section_poisson}
Considering the discrete numerical nature of the response data, an additional candidate model for the specific analysis, is the Poisson regression, as also suggested by a referee.
In our context, the Poisson regression model is given by:

\begin{equation}
\text{Pts}_i \sim \text{Poisson}(\xi_i) \quad 
\label{poissonModel}
\end{equation}
where $\text{Pts}_i$ is the number of points scored in possession $i$, $\xi_i$ is the expected number of points scored in possession $i$,  and $\log(\xi_i)$ has the same form as $\mu_i$ in \eqref{normal}. 
Regarding the regularization implementation of the Poisson model, the lasso is used as in the previous implementations. 
Using lasso for Poisson regression models is common in football; see for example in \cite{groll2015} who implemented $L_1$-penalized Poisson regression to predict outcomes in international soccer tournaments.

\begin{figure}
    \centering
    \includegraphics[width=0.6\linewidth]{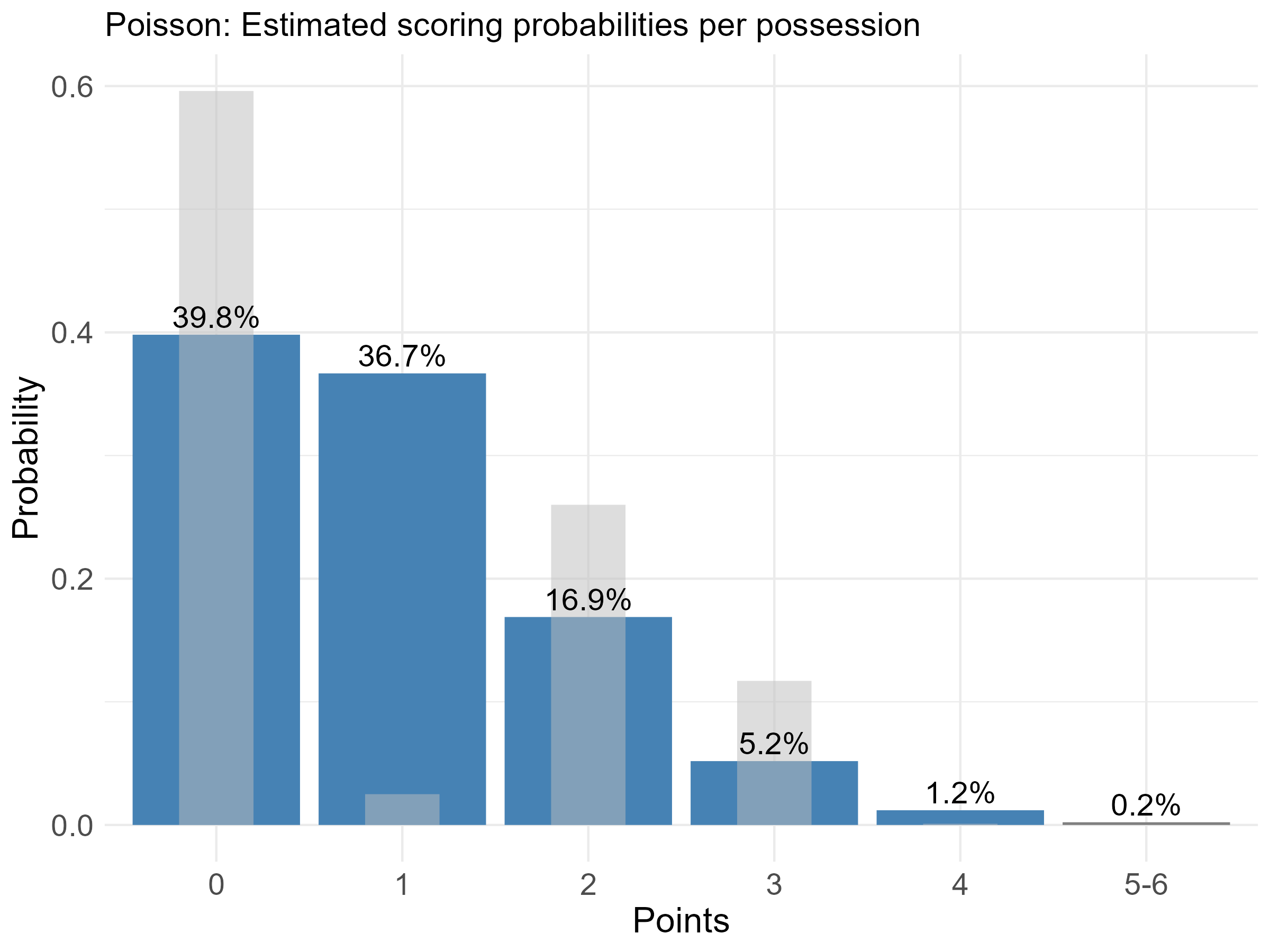}
    \caption{Estimated probabilities for each scoring category obtained using the Poisson model. The narrow, light gray bars represent the empirical frequencies of points per possession (in the category of 4, 5 and 6 points the frequency is quite low).}
    \label{fig:poisson_probs}
\end{figure}

An important limitation of the Poisson model is that the response variable in our case study (i.e., the number of points per possession) is not defined over the entire set of natural numbers, as would be expected for a Poisson-distributed variable. Additionally, a key distinction between the two models lies in their interpretations: the Poisson model focuses on modeling the overall average number of points scored per possession, which indirectly determines the scoring probabilities, while the multinomial model directly estimates each player's effect on individual scoring probabilities.
As a consequence, in a Poisson model, an increased average contribution leads to higher probabilities of scoring two or three points. However, this is not typical of how points are distributed in basketball. Figure \ref{fig:poisson_probs} illustrates a comparison between the observed and Poisson-fitted probabilities in our dataset. The probabilities of the Poisson model shows a decreasing pattern, whereas the observed probabilities follow a non-decreasing pattern, with the highest probability for no points, followed by two points, three points, and lastly, one point (which is very low).  
As a result, the Poisson model does not appear to be suitable for modelling the number of points per possession  since it does not capture the actual scoring pattern in basketball.



Finally, when considering external validation Criteria 1–4, the Poisson model performs similarly to the normal RAPMs, being slightly worse in Criteria 1–3 but better in 4 out of 6 box scores used for Criterion 4. However, when compared to the multinomial model, the Poisson model performs worse in relation to the box scores for Criterion 4: the multinomial model outperforms the Poisson in 4 out of 6 criteria, while the Poisson is better in only one (blocks), and both models perform equally on offensive rebounds. 

To conclude, the Poisson model is not compatible with the possible outcomes for the number of points per possession modeled in this work. Its final probability distribution fails to capture the non-decreasing pattern of probabilities observed in basketball, and the Poisson model performs worse than the multinomial model in external validation criteria based on box scores. 
Thus we cannot propose its use for estimating players contributions in per-possession play-by-play data.

\subsection{Ordered Multinomial Model implementation}
\label{section_ordinal}

The ordered multinomial regression serves as an alternative to the multinomial model, treating the response variable as ordinal. However, the ordered multinomial regression relies on the important assumption of proportional odds. 
This assumption states that the relationship between each type of points scored per 
possession\footnote{measured by the differences in log-odds of cumulative scoring probabilities for subsequent types} 
remains the same across different specifications of the team lineups.  
For instance, the log-odds of scoring two points or less, minus the log-odds of scoring one point or no points, should be constant regardless of the team lineups.

\begin{figure}
    \centering
    \includegraphics[width=0.7\linewidth]{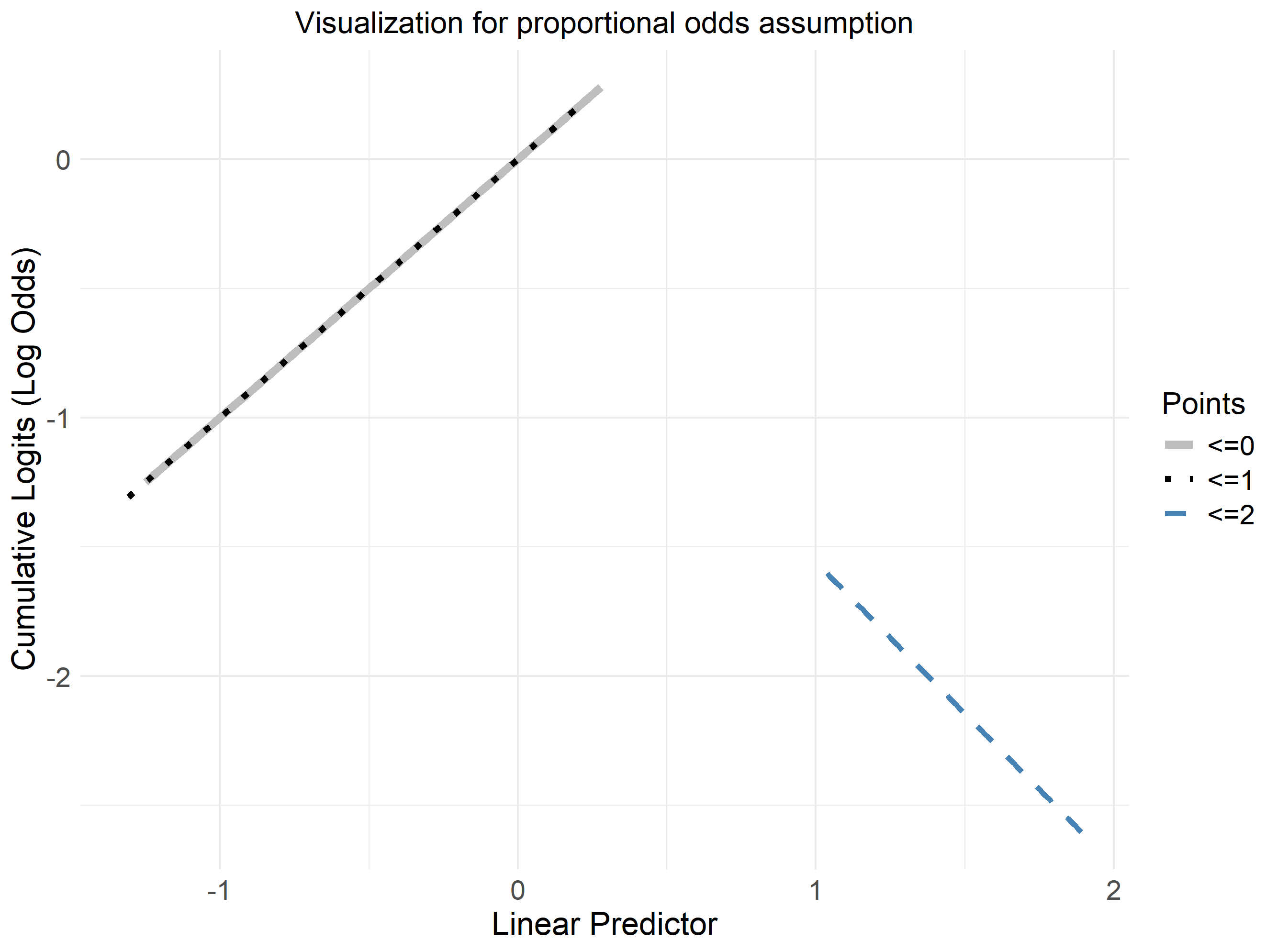}
    \caption{Log-odds of cumulative probabilities across the linear predictor for each scoring category.}
    \label{fig:ordinal_lasso_propOdds}
\end{figure}

However, this assumption is clearly violated in per-possession basketball data, as each player's efficiency varies across different types of points. Some players are exceptional at three-pointers, while others are more effective close to the basket, making them more accurate at scoring two-pointers. 
This inconsistency is evident in Figure \ref{fig:ordinal_lasso_propOdds}, where 
the variation of the players' coefficients 
across different scoring categories indicates a violation of the proportional odds assumption, suggesting that players' effects are not consistent across the levels of the response variable. 
The violation of the proportional odds assumption is further confirmed by the Brant test (p-value $<$ 0.001) 
implemented for the non-regularized ordinal logistic model. 
Finally, the ordinal model appears to have very close performance with regard to the external validation Criteria 1--4.

\section{A New Improved Evaluation Metric: Weighted Expected Points (wEPTS)} 
\label{sec4_wEPTS}

\subsection{The metric}
\label{section_define_WEPTS}
In this section, we proceed by introducing an improved weighted version of EPTS which takes into consideration the participation of each player throughout the season. 
The goal of this evaluation metric is to quantify the expected points per possession for the team of the player under study. We will evaluate two different lineup configurations: one lineup including the player of interest and a second one without the player of interest. All other players on the court are assumed to belong to the reference category.  The corresponding expected points of these two line-ups will be weighted according to the proportion of possessions in which the player of interest participated. 

Hence,   the weighted Expected Points (wEPTS) per possession is defined as  the expected team points per possession for the team of $k$  player when all other players are from the reference group with the zero-lasso RAPMs and it is given by 
\begin{eqnarray} 
wEPTS^{r}_{k} 
&=& 
E(Pts_i |  {\bf X}^r_{i\setminus k} = {\bf 0}, {\bf X}^{\overline{r}}_{i} = {\bf 0}, T_i=t_k ) \nonumber \\[1em]
&=& 
~~~~ P(X_{ik}^r \neq 0 | T_i=t_k) 
E(Pts_i | X_{ik}^r \neq 0, {\bf X}^r_{i\setminus k} = {\bf 0}, {\bf X}^{\overline{r}}_{i} = {\bf 0})  \nonumber \\ 
&& +P(X_{ik}^r = 0 | T_i=t_k) 
E(Pts_i | X_{ik}^r = 0, {\bf X}^r_{i\setminus k} = {\bf 0}, {\bf X}^{\overline{r}}_{i} = {\bf 0} ) 
\nonumber \\[1em]
&=&  W_k^r \, EPTS_k^{r} + (1-W_k^r) \, EPTS_0^{r} \label{wEPTS} 
\end{eqnarray}
where $r \in \{o, d\}; ~~\overline{r} =  \{o, d\} \setminus r$ . Moreover, the weight $W^r_k$  is estimated by 
$$ 
\widehat{W}^r_k= \frac{n^{r}_{k} }{n^{r}_{t_k}} 
=\dfrac{\sum_{i=1}^n {\cal I}(X_{ik}^r\neq 0)}{ \sum_{i=1}^n {\cal I}(T_i^r=t_k)}, 
$$
and ${\cal I}(A)$ is the indicator function taking value one if $A$ is true and zero otherwise; 
$T_i$  is the team in offence/defence in possession $i$, 
$t_k$ is the team of player $k$, 
$X^{r}_{ik}$ is the binary dummy variable for player $k$ taking zero-one values in offensive ratings and minus one-zero in the defensive ratings, 
$({\bf X}^r_{i\setminus k} = {\bf 0}, {\bf X}^{\overline{r}}_{i} = {\bf 0} )$ implies that all other players except for the player of interest are set equal to players belonging to the reference group, 
$n^{r}_{t_k}$ is the number of possessions that the team of player $k$ is in offence or defence (depending on $r$),  
$n^{r}_{k}$is the number of possessions that the player $k$ is included in the playing lineup of the team in offence or defence. 
Finally, $EPTS_k^{r}$ are the expected points of player $k$ as defined \eqref{epts} while $EPTS_{0}^r$ are the expected points of the reference lineup, given by 
\begin{equation}
EPTS_{0}^r 
= \dfrac{e^{b_{01}} + 2  e^{b_{02}} + 3.01  e^{b_{03}} }
{1+e^{b_{01}} +  e^{b_{02}} + e^{b_{03}}}
\mbox{ for} ~r \in \{o, d\}.
\label{epts0}
\end{equation}

The wEPTS rating has a similar interpretation to EPTS. 
The primary distinction is that we now consider the conditional expectation of the points for a reference lineup with and without the player of interest weighted by the proportion of possessions in which the player of interest participates on his team. 
Overall, this index acts as a shrinkage method on EPTS shrinking them towards the expected points of the reference group for players with minimal playing time.  
On the other hand, if we consider the case of a player who participates in all possessions of his team (which is not a realistic case in practice) then the $wEPTS$ will become equal to the $EPTS$. 
Finally, the new index only requires minimal extra computations since we only need 
the proportion of possessions that each player participates in his team and to calculate 
the expected points of the reference group  $EPTS_0^r$ given by \eqref{epts0}  which is nevertheless, directly available from the lasso coefficients. 
 
As referee suggested, the weighted EPTS  resembles the method for calculating expected values in the mixture framework. However, this literature discusses the interpretation of expected values computed as a weighted mean of zeros and positive values. In fact, the expected values for the quantitative outcome consistently show a downward shrinkage. 
In our context, shrinkage to zero is desirable because we aim to estimate the overall contribution of players to their team. This contribution should be influenced by the amount of time a player spends in the game. Players with limited playing time (e.g., LTPs) are unlikely to have a significant impact on their team, even if their performance during play is exceptional. Therefore, EPTS and wEPTS can be used complementary: EPTS can identify a limited number of players with “positive” performance even with low playing time, while wEPTS measures overall contribution, which accounts for playing time. By using wEPTS—a weighted mean of zero (when the player is not playing) and their mean points (when playing)—we provide a robust and meaningful measure of player contribution that addresses the LTP issue.

\subsection{Weighted EPTS (wEPTS) through validation criteria}
\label{section_WEPTS_comparison} 
We proceed with the comparison of wEPTS with the EPTS rankings and the RAPMs from the lasso-normal approach with respect to the external validation criteria we have introduced in Section  \ref{ext_val_criteria}.

Tables~\ref{tab:weighted_criteria} and ~\ref{tab:weighted_boxscore_criteria} demonstrate the superiority of the weighted EPTS across various criteria. 
Furthermore, when examining the list of players with the highest EPTS and wEPTS ratings, we reveal that the average playing time per game is 26.3 and 29.7 minutes, respectively. 
Furthermore, among these top-rated players, the average points scored per game are 15.1 and 17.8 for EPTS and wEPTS, respectively. Concurrently, when considering players with the lowest offensive EPTS ratings, they demonstrate considerably higher points, on average, than the corresponding bottom list players of offensive wEPTS, averaging 16.6 and 9.6 points per game, respectively.

\begin{table}[h!]
    \caption{Comparison of External Validation Metrics for Criteria 1--3 for lasso Ratings (Normal, Multinomial and weighted Multinomial).}
    \label{tab:weighted_criteria}
    \centering
    \begin{tabular}{|>{\raggedright\arraybackslash}p{3cm}|c c c c|}
    \hline
    \textbf{Rating (Model)} & Criterion \#1 & Criterion \#2 & Criterion \#3a & Criterion \#3b\\ [0.5ex]
    \hline\hline
    RAPM (Normal) & 40\% & 3\% & 55\% & 23\%\\
    EPTS (Multinomial) & 47\% & 9\% & 47\% & 20\%\\
    wEPTS (Multinomial) & \textbf{67\%} & \textbf{2\%} & \textbf{74\%} & 45\%\\
    \hline
    \multicolumn{5}{p{11.5cm}}{\footnotesize \it Criterion \#1: All-NBA teams, Criterion \#2: Low-time players in 100 top-RAPM, Criterion \#3a: Starters in 100 top-RAPM and Criterion \#3b: Starters in 100 bottom-RAPM; \textbf{Bold} indicates the maximum value by column/Criterion} \\
    \end{tabular}
\end{table}

\begin{table}[h!]
    \caption{
Comparison of External Validation Criterion 4 for lasso Ratings (Normal, Multinomial and weighted Multinomial).}
    \label{tab:weighted_boxscore_criteria}
    \centering
    \begin{tabular}{|>{\raggedright\arraybackslash}p{3cm}|c c c c c c|}
    \hline
    \textbf{Model} & PTS & AST & OREB & DREB & STL & BLK\\ [0.5ex]
    \hline\hline
    RAPM (Normal) & 42\% & 26\% & 14\% & 8\% & 14\% & 12\%\\
    EPTS (Multinomial) & 38\% & 28\% & \textbf{16\%} & 14\% & 20\% & 14\%\\
    wEPTS (Multinomial) & \textbf{44\%} & \textbf{36\%} & 14\% & \textbf{22\%} & \textbf{28\%} & \textbf{22\%}\\
    \hline
    \multicolumn{7}{p{10cm}}{\footnotesize \it PTS: points scored per game, AST: assists per game, OREB: offensive rebounds per game, DREB: defensive rebounds per game, BLK: blocks per game, STL: steals per game; \textbf{Bold} indicates the maximum value by column/box-score} \\
    \end{tabular}
\end{table}    

Although wEPTS demonstrates clear superiority across (nearly) all selected validation criteria, a surprising result is found concerning Criterion 3b in Table~\ref{tab:weighted_criteria}. 
It becomes apparent that a higher proportion of starters appear in the list of players with the lowest wEPTS rankings compared to the corresponding list compiled by using EPTS ($45\%$ and $20\%$ respectively).
However,  none of the starters was found to be in the bottom list of both offensive and defensive ratings. 
This suggests that such starters are ranked in the bottom 50 list of offensive ratings due to their primary defensive role within the team, wherein their defensive contribution is considerably higher than their offensive performance. Conversely, all starters who appeared in the bottom 50 list of defensive ratings exhibit substantially higher offensive contributions, potentially indicative of their offensive responsibilities within the team.

\subsection{Lasso vs. After-Lasso Ratings}
\label{section_lasso_vs_after-lasso}
The RAPM and EPTS ratings presented in the previous sections are based on biased ``shrunk'' estimates of model coefficients which quantify the contribution of each player. 
This section explores the impact of bias on RAPM and EPTS ratings. 
We compare two approaches: (a) using directly biased coefficients derived using lasso and (b) using the unbiased (MLE) coefficients, after removing players whose coefficients were flagged as zero by lasso.
For the latter case we will use the conventional name ``after-lasso'' ratings. 

Tables~\ref{tab:weighted_criteria_afterLasso} and \ref{tab:weighted_boxscore_criteria_afterLasso} 
present the performance of the two approaches (lasso and after-lasso) for the selected validation criteria. 
From these tables, there is no clear winner between the two approaches 
and there are minor differences in performance between the lasso coefficients and the after-lasso coefficients, particularly for the proposed wEPTS ratings. Given that the two approaches are similar in performance and the additional computational burden required to refit the model for after-lasso ratings, we recommend directly using the lasso coefficients for the calculation of wEPTS.

\begin{table}[h!]
    \caption{Comparison of validation Criteria 1--3 between lasso and after-lasso ratings for the Normal RAPM, the Multinomial EPTS and the Multinomial wEPTS.}
    \label{tab:weighted_criteria_afterLasso}
    \centering
    \begin{tabular}{|>{\raggedright\arraybackslash}l l | c c c|}
    \hline
    \textbf{Model} &&&& \\ 
    \textbf{(Rating)} &  \textbf{Method} & Criterion \#1 & Criterion \#2 & Criterion \#3a\\ [0.5ex]
    \hline\hline
    \small{Normal} &  \small{Lasso} & 40\% & 3\% & 55\%\\
    \rowcolor{gray!20} (\small{RAPM}) &\small{After-lasso} & 33\% & 6\% & 50\%\\
    \hline 
    \small{Multinomial}  & \small{Lasso} & 47\% & 9\% & 47\%\\
    \rowcolor{gray!20} (\small{EPTS}) & \small{After-lasso} & 33\% & 10\% & 40\%\\
    \hline 
    \small{Multinomial} & \small{Lasso} & 67\% & 2\% & 74\%\\
    \rowcolor{gray!20} (\small{wEPTS})& \small{After-lasso} & 60\% & 2\% & 73\%\\
    \hline
    \multicolumn{5}{p{10cm}}{\footnotesize \it Criterion \#1: All-NBA teams, Criterion \#2: Low-time players in 100 top-RAPM, and Criterion \#3a: Starters in 100 top-RAPM} \\
    \end{tabular}
\end{table}

\begin{table}[h!]
    \caption{Comparison of validation Criterion 4 between lasso and after-lasso ratings for the Normal RAPM, the Multinomial EPTS and  the Multinomial wEPTS.}
    \label{tab:weighted_boxscore_criteria_afterLasso}
    \centering
    \begin{tabular}{|>{\raggedright\arraybackslash}l l | c c c c c c|}
    \hline
    \textbf{Model} &&&&&&& \\ 
    \textbf{(Rating)} & \textbf{Method} & PTS & AST & OREB & DREB & STL & BLK \\ [0.5ex]
    \hline\hline
    \small{Normal} &  \small{Lasso} & 42\% & 26\% & \textbf{14\%} & 8\% & 14\% & 12\%\\
    \rowcolor{gray!20} (\small{RAPM}) &\small{After-lasso} & \textbf{44\%} & \textbf{30\%} & 12\% & \textbf{20\%} & \textbf{22\%} & 12\%\\
    \hline 
    \small{Multinomial}  & \small{Lasso} & \textbf{38\%} & 28\% & \textbf{16\%} & 14\% & 20\% & 14\%\\
    \rowcolor{gray!20} (\small{EPTS}) & \small{After-lasso} & 36\% & \textbf{30\%} & 12\% & 14\% & \textbf{22\%} & 14\%\\
    \hline 
    \small{Multinomial} & \small{Lasso} & 44\% & 36\% & 14\% & 22\% & 28\% & \textbf{22\%}\\
    \rowcolor{gray!20} (\small{wEPTS})& \small{After-lasso} & \textbf{50\%} & \textbf{38\%} & \textbf{16\%} & 22\% & \textbf{34\%} & 20\%\\
    \hline
   \multicolumn{8}{p{10.5cm}}{\footnotesize \it PTS: points scored per game, AST: assists per game, OREB: offensive rebounds per game, DREB: defensive rebounds per game, BLK: blocks per game, STL: steals per game; \textbf{Bold} indicates the maximum value by method} \\
    \end{tabular}
\end{table}

\subsection{The Effect of Low Time Players}
\label{section_effect_of_ltps}
In Section \ref{section_Filtering_ltps}, the implementation of a cut-off threshold for minutes played by the studied players is examined in relation to the higher performance of the regularized Gaussian models with respect to players ranking. Based on these improved results, we subsequently applied the logistic multinomial model to the ``filtered" dataset, excluding low-time players from the analysis. Nevertheless, it would be worthwhile to examine the proposed methodology by considering all players in the analysis. 

\begin{figure}
    \centering
    \includegraphics[width=0.7\linewidth]{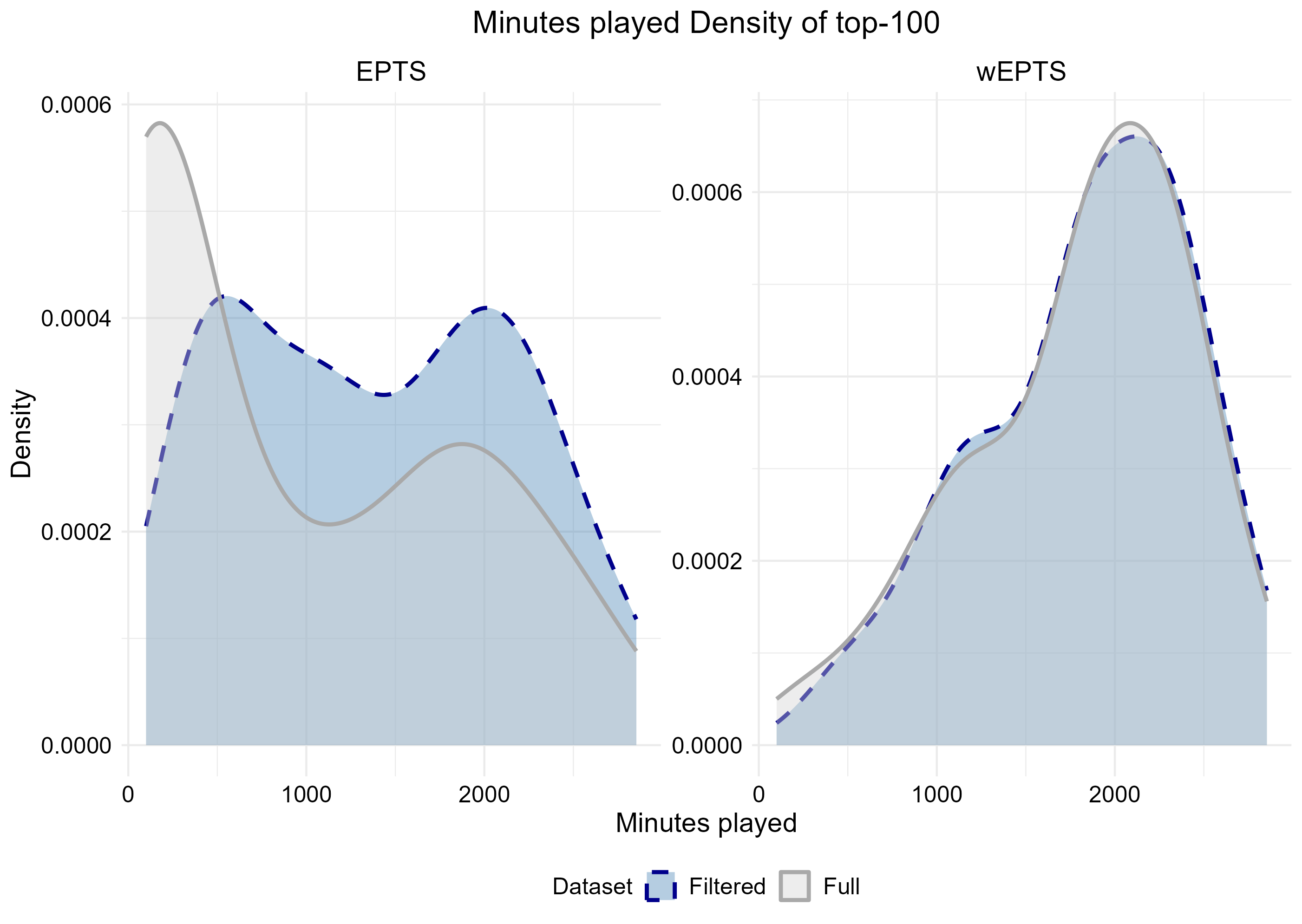}
    \caption{Density of top-100 contributed players playing time based on the Multinomial approach followed by including (Full dataset) or excluding low-time players (Filtered dataset).}
    \label{fig:density_MP_top100_full_dataset}
\end{figure}

In fact, we expect coming across similar issues regarding the low-time players (LTPs) during the development of the EPTS; however, we do not expect such issues to arise with the wEPTS. Our intuition is confirmed, as shown in Figure \ref{fig:density_MP_top100_full_dataset}, where the distribution of the top EPTS contributed players is right skewed when LTPs are incorporating in the analysis, and more flat when they are excluded; additionally, LTPs (less than 200 minutes in the season) constitute 38\% of top 100 in the full data analysis, indicating that LTPs have a significant impact on the results. As for the wEPTS, the corresponding distribution is similar for both implementations. 
Hence, wEPTS appears to address the issue of LTPs, as the distribution of the top 100 remains robust despite the presence of LTPs.

\section{Concluding remarks}
\label{sec5_discussion}

\subsection{Conclusion}
Following the literature, we initially implemented ridge regression on the full possession dataset for the 2021-2022 NBA season.
The RAPM (Regularized Adjusted Plus-Minus) model is generally applied to data aggregated over shifts, where the composition of the team on the court remains constant. This aggregation lends itself to the use of Gaussian models. However, applying a Gaussian model directly to disaggregated data presents significant challenges in terms of justification. Implementing a linear model in this context can lead to a substantial oversimplification and may introduce bias, as has been observed.
Specifically, this approach was susceptible to the influence of low-time players (LTPs), whose limited playing time inflated their estimated contributions, despite the shrinkage. 
To address this issue, we have investigated two different approaches: 
First, we have implemented lasso instead of ridge and, second, we have removed low-time players from RAPM estimation. 

For the first approach, lasso regression promotes sparsity by setting specific coefficients equal to zero. 
This will be effective for less influential variables (players). 
This characteristic of lasso resulted in RAPM estimates with a clearer distinction between well-performed, average and low-performed players. 
Concerning LTPs, the problem was mitigated (but not diminished) by shrinking the coefficients of such players to zero 
(i.e. to the group of ``average'' players). 
On the other hand, the exclusion of LTPs  (defined as less than 200 minutes for the season)  also improved the performance of the RAPM ratings.
For this reason, we have proceeded by combining the two strategies (i.e. implementing lasso without LTPs). 

The lasso model demonstrated better performance when evaluated using external validation criteria (Section~\ref{ext_val_criteria}). However, a key characteristic of lasso, while advantageous in some respects, is its tendency to set coefficients to zero, effectively omitting players with doubtful contributions. 
However, in the normal model, lasso regression resulted in the shrinkage of approximately $70\%$ of the player coefficients to zero. Consequently, only the remaining $30\%$ of players have been assigned with non-zero RAPM ratings.

Given the discrete nature of the response variable, a logistic regression model is the logical next step in our analysis. Initially, a binary classification model was fitted, predicting the outcome of scoring versus not scoring. 
Regularized logistic regression was implemented, achieving an accuracy of approximately $55\%$. 
However, this approach provides a simplified solution to our problem, neglecting the actual number of points scored on each possession.

Despite the moderate accuracy, an intriguing finding emerged from the relationship between the ridge binomial logistic and ridge normal RAPMs. We observed a strong linear relationship between the ratings generated by these two models. 
This finding provides a compelling justification for using the Normal RAPMs, despite their initial limitation of being derived from a model which is not ideally suited for the use of the number of points as a response. 
The fitted linear relationship allows us to convert normal RAPMs into logistic regression RAPMs, which are obtained through a methodologically sound approach. 
This conversion essentially transforms the normal RAPM scores into a framework more appropriate for the binary outcome (scoring vs. not scoring), enhancing the interpretability of standard RAPM player performance metrics.

Finally, the multinomial model --- which is more appropriate for this type of response --- was applied to the number of points per possession. To circumvent computational issues and to introduce greater flexibility in terms of which lasso coefficients were set to zero, we employed three distinct binomial implementations.
Moreover, the multinomial model has the benefit of providing more specific information about each player's contribution to the various scoring formats (one, two, or three points). Higher-ranked players in terms of three-point shooting might be chosen to play, for instance, if a team requires a three-point shot at a critical moment of the game. It is not about players with great shooting percentages; rather, it's about players who boost their team's chances of scoring a particular type of points. 
Such players have a specific role in the team, as for example a center with ball-passing ability or a player who drives to the paint with ease and passes the ball to the desired player for a shot.

Considering insights from the multinomial model, we introduce a novel RAPM metric: the Expected Points (EPTS). This metric is derived from a correctly specified model for the number of points scored. To further refine the analysis, we propose a weighted version of EPTS, denoted as wEPTS. 
This extension incorporates player participation within each team's possessions, effectively handling the issue of inflated contributions observed with low-time players. 
In summary, EPTS is based on a statistically sound foundation for evaluating player performance based on expected points scored. Moreover, our evaluation using established validation criteria demonstrates that wEPTS outperforms all other approaches considered in this study.

Considering our reviewers' suggestions, and given the discrete or ordinal nature of points per possession (our response), two additional models were examined; a Poisson and an ordered multinomial regression model. 
The results confirmed our initial intuition that these models were not appropriate for our response data (i.e. the points per possession). Firstly, the Poisson model incorrectly assumes that the probability of scoring one point is higher than two and three, which is not generally true in basketball. Secondly, the proportional odds assumption for the ordered multinomial model is violated, since the effect of a player on the odds of one scoring category compared to all higher categories is not constant across all thresholds of the ordinal outcome.

To conclude, in this study, we have considered a variety of issues concerning the estimation of RAPM player ratings from possession-based NBA data. The main contributions of this paper are: 
(a) we suggest the use of lasso instead of ridge and demonstrate its superiority in terms of evaluations; 
(b) we offer a clear interpretation of commonly used normal ridge RAPMs through  logistic regression coefficients; 
and finally, 
(c) we introduce novel RAPM metrics: Expected Points Scored (EPTS) and its weighted version (wEPTS) based on a multinomial logistic regression model, which is the statistically appropriate approach for modelling the points scored per possession.
The proposed wEPTS has four clear advantages: 
First, it outperforms all other approaches evaluated in this study, including the standard ridge normal RAPMs. 
Secondly, it effectively addresses the challenge of low-time players, almost eliminating their inflated contributions. 
Third, it is based on a foundationally appropriate statistical model and, four,  it can separately evaluate and consider the contribution of each player based on the different type of scoring points that each player contributes.

\subsection{Discussion and future research}
Due to the nature of basketball, where players interact in offence and defence in order to perform better as a team, analyzing the impact of lineups, rather than solely individual players, is quite important. The Regularized Adjusted Plus-Minus (RAPM) evaluates players performance, while their teammates and opponents are considered in its development. 
The weighted Expected Points per Possession (wEPTS) we propose in this work has a similar value, but at the same time, it addresses issues found in Gaussian RAPMs, such as accounting for the discrete nature of the response variable, the varying contributions of players to different types of points, and the effect of LTPs.

The proposed rating of weighted EPTS (wEPTS) appears with approximately 78\% agreement (while 63\% corresponds to EPTS) between the lineups determined for each team, considering the highest offensive and defensive contributed players per position, and the two most frequently used (most-played) lineups per team in the season. Although this result seems surprising, it does not imply that simply combining individual player contributions would lead to a useful evaluation or construction of more effective lineups, since our approach is designed to capture the individual players' effects.

In the literature, the effect of lineups is examined using methodologies similar to those applied to individual player performance ratings, replacing individual players with lineups in the analysis. 
This simplified solution could be applied to estimate the lineups effect through a multinomial logistic model. 

Regarding this, \cite{grassetti2020} introduced an alternative way to estimate lineups effect, including both lineup and players effects in the model specification; suggesting that the first can be seen as a type of higher-order interaction among players. In fact, each player demonstrates a varying level of impact depending on the lineup in which he is included. 
Similar ideas could be introduced in our proposed methodology, by incorporating the effect of lineups or similar adjustments in order to capture or at least consider the interaction. 

In this context, ``interaction" is the key, since investigating and detecting possible patterns in the court is more than desirable. Within a lineup, each player has a specific role that complements the roles of their teammates. Therefore, considering the abilities and tendencies of each player, coaches and staff try to optimize their roster under dynamic situations. This is the main challenge of such analyses; dynamic situations arise from a variety of factors. 
Thus, an analysis grounded in a team's in-game habits and style of play may prove to be both promising and highly valuable, but also challenging. 

Considering the disaggregated data that were used, some relevant covariates may be useful. For instance,  \cite{grassetti2020} included the game period and the type of shots (close-range, mid-range, three-pointers) in their proposed model, while they used some boxscore statistics to define the response variable based on the outcome of a possession. Following these ideas, some candidate covariates would be the time till the end of the game (the garbage time could be considered as an extra indicator), the score difference at each possession-event (or a variation according to how a field goal made decreases or increases the score difference of a team), the time of an attempted shot (made or not) or a turnover during offence (fast break, secondary transition, and, set), which can be connected to a team's style of play. 

Moreover, the competition level and structure of each basketball league vary. For instance, an average NBA team has the opportunity to compete for postseason qualification even in the most recent regular season games, but an average Euroleague team needs to try harder during the regular season in order to proceed to the playoffs.
This aspect could be reflected through the importance of a game within the season and mathematically converted into an appropriate, and dynamic weight.

Thus, in order to improve the accuracy of the players' performance ratings, the proposed approach can include the extra variables described in the previous two paragraphs. 

Moreover, the suggested methodology might incorporate different shrinkage techniques. 
Avoiding underestimation or overestimation of players contributions, especially in the case of low-time players, is of main interest. In recent years, there has been a great enlargement in shrinkage literature, especially through Bayesian techniques. Global-local shrinkage methods, like spike-and-slab (introduced by \citealp{ssvs_mcculloch_1993}) and horseshoe (introduced by \citealp{carvalho_2010}) prior assuming on players effect seem suitable and promising for our case study. These two approaches are similar; in the first, the local shrinkage parameter is typically fixed, whereas in the second, it is distributed using a half-Cauchy distribution. As an additional step in our suggested process, we can take player participation into account for this parameter. Alternatively, we may choose not to apply any of these shrinkage priors and instead set an ``informative'' prior based on the playing time of each player to account for their individual effect.   
Even while this sounds interesting, we need to take certain computational issues into consideration.

\section*{Acknowledgement}
This research is funded by the Scholarship and Research Program of the Department of Statistics of Athens University of Economics and Business and by the Institute of Statistical Research Analysis and Documentation (ISTAER).
\bibliography{sn-bibliography}

\end{document}